\documentclass{article}
\usepackage{amsmath}
\usepackage{amssymb}
\usepackage{multirow}
\usepackage{multicol}
\usepackage{dblfloatfix}
\usepackage{setspace}
\usepackage{lineno}
\usepackage{subfig}
\usepackage{lscape}
\usepackage{changepage}
\usepackage{psfrag}
\usepackage{color}
\usepackage{rotating}
\usepackage{tablefootnote}
\usepackage{arydshln}
\usepackage{gensymb}
\usepackage{url}
\usepackage{geometry}
\usepackage{natbib}

\newcommand{\bI}{ {\boldsymbol I} }

\newcommand{\bx}{ {\boldsymbol x} }

\newcommand{\by}{ {\boldsymbol y} }

\newcommand{\bSigma}{ {\boldsymbol \Sigma} }

\newcommand{\bzero}{ {\boldsymbol 0} }
\newcommand{\bones}{ {\boldsymbol 1} }

\newcommand{\N}{\mathcal{N}}

\newcommand{\specialcell}[2][c]{
  \begin{tabular}[#1]{@{}c@{}}#2\end{tabular}}

\providecommand{\keywords}[1]{\small \textbf{\textit{Keywords---}} #1}

\title{Remote sensing to reduce the effects of spatial autocorrelation on design-based inference for forest inventory using systematic samples}
\date{}

\author{Chad Babcock$^{1}$\thanks{Corresponding author (babcoc76@msu.edu)}, Andrew O. Finley$^{1}$, Timothy G. Gregoire$^{2}$, Hans-Erik Andersen$^{3}$\\[.5em]
\small{$^{1}$Michigan State University, Department of Forestry, East Lansing, MI 48824, USA}\\
\small{$^{2}$Yale School of Forestry \& Environmental Studies, New Haven, CT 06511, USA}\\
\small{$^{3}$USDA Forest Service, Pacific Northwest Research Station, Seattle, WA 98195, USA}}
 
\begin{document}
\maketitle
\begin{abstract}
  Systematic sampling is often used to select plot locations for forest inventory estimation. However, it is not possible to derive a design-unbiased variance estimator for a systematic sample using one random start. As a result, many forest inventory analysts resort to applying variance estimators that are design-unbiased following simple random sampling to their systematic samples even though this typically leads to conservative estimates of error. We explore the influence of spatial autocorrelation on variance estimation when systematic sampling is employed using repeated sampling. We generate a sequence of 1000 synthetic populations with increasing spatial autocorrelation and repeatedly sample from each to examine how the performance of estimators change as spatial autocorrelation changes. We also repeatedly sample from a tree census plot in Harvard Forest, Massachusetts and examine the performance of similar estimators. Results indicate that applying variance estimators that are unbiased following simple random sampling to systematic samples from populations exhibiting stronger spatial autocorrelation tend to be more conservative. We also find that incorporating ancillary wall-to-wall covariates, e.g., remote sensing data, using generalized regression estimators can reduce variance over-estimation by explaining some or all of the spatial structure in the population.
\end{abstract}

\keywords{design-based inference, forest inventory, remote sensing, spatial autocorrelation, systematic sampling}

\section{Introduction}
Systematic sampling is standard practice in many forest inventory settings, ranging from stand-level timber cruising to national forest inventory programs \citep{avery2002, gillis2005}. Systematic sampling is often easier to implement and results in more precise inventory estimates than simple random sampling with the same number of field observations \citep[p. 205]{cochran1977}. However, in design-based inference, there is no way to derive a design-unbiased variance estimator using a single systematic sample \citep[p. 27]{mandallaz2008}. Often, estimators that are design-unbiased following simple random sampling are used in their stead \citep[p. 55]{gregoire2008}. When this type of estimator is used, the potential efficiency gains of systematic sampling are not reflected in the resulting variance estimate, i.e., this variance estimator can be positively biased when used with systematic samples. It is understood that the underlying structure of the population, e.g., the degree of ordering among the population units, effects how positively bias the estimator will be. \citet[pp. 78-82]{sarndal1992} provides an excellent example of the effects of population ordering on systematic sampling efficiency.

Systematic sampling with one random start can be viewed as cluster sampling where only one cluster is drawn \citep[ch. 12]{thompson2012}. In traditional cluster sampling where groups (clusters) of observations are made in close proximity to each other, within cluster variability is often less than between cluster variability, making cluster sampling less efficient than simple random sampling given the same number of observations. The opposite can be true for systematic sampling where sample observations are forced to be far apart. In a population exhibiting positive spatial autocorrelation, variability within systematic samples can be greater than variability between systematic samples. Cluster sampling estimators can account for potential variance estimator bias, e.g., by incorporating a \emph{design effect} term \citep{gambino2009, kish1995}. The calculation of which requires more than one cluster. So even if one views systematic sampling with one random start as a special case of cluster sampling, they are still out of luck for calculating a \emph{design effect}. \citet[p. 185-186]{iles2003} provides an example variance estimator calculation that is design-unbiased following systematic sampling using multiple random starts.

The advent of inexpensive wall-to-wall remote sensing information relevant to forest structure and composition, e.g., lidar and multispectral data, is encouraging the use of model-assisted estimators for forest inventory to improve precision \citep{mcroberts2013,naesset2011,saarela2015}. In design-based model-assisted inference, covariates derived from remotely sensed data are commonly related to forest inventory measurements via regression and the inventory attribute of interest is predicted at locations where only the remote sensing covariate is observed. The mean of the predictions (bias-corrected using model residuals) and residuals can then be used to estimate the population mean and associated estimator variance, respectively. This is commonly known as a generalized regression (GREG) estimator \citep{gregoire2008}.

In addition to improving precision of estimation, we posit that wall-to-wall remotely sensed covariates can reduce variance estimator bias following systematic sampling by using GREG estimators. Variance estimator bias following systematic sampling is, in part, affected by spatial autocorrelation in the population \citep{flores2003}. Using covariates strongly correlated with the population values being averaged using a GREG estimator can account for spatial autocorrelation, potentially leading to reductions in variance estimator bias following systematic sampling.

We aim to assess the influence of spatial autocorrelation on variance estimators following systematic and simple random sampling. We simulate a series of 1000 synthetic populations with increasing spatial structure. Using repeated sampling (simple random and systematic), we compare the mean of design-based Hortwitz-Thompson (H-T) and GREG variance estimates for each population to empirical estimator variance. In addition, we conduct a similar repeated sampling experiment on a 35 hectare (ha) stem map located in Harvard Forest, Massachusetts where lidar and hyperspectral remote sensing data were collected. We use the results from the synthetic populations and Harvard Forest analyses to examine: 1) how variance estimator performance changes with changes in spatial autocorrelation following systematic and simple random sampling; 2) how assisting model covariates can reduce variance estimator bias by accounting for spatial autocorrelation.

\section{Synthetic Populations Analysis Methods}\label{synth-description}

\begin{figure}[!h]
\centering
\includegraphics[width = \textwidth]{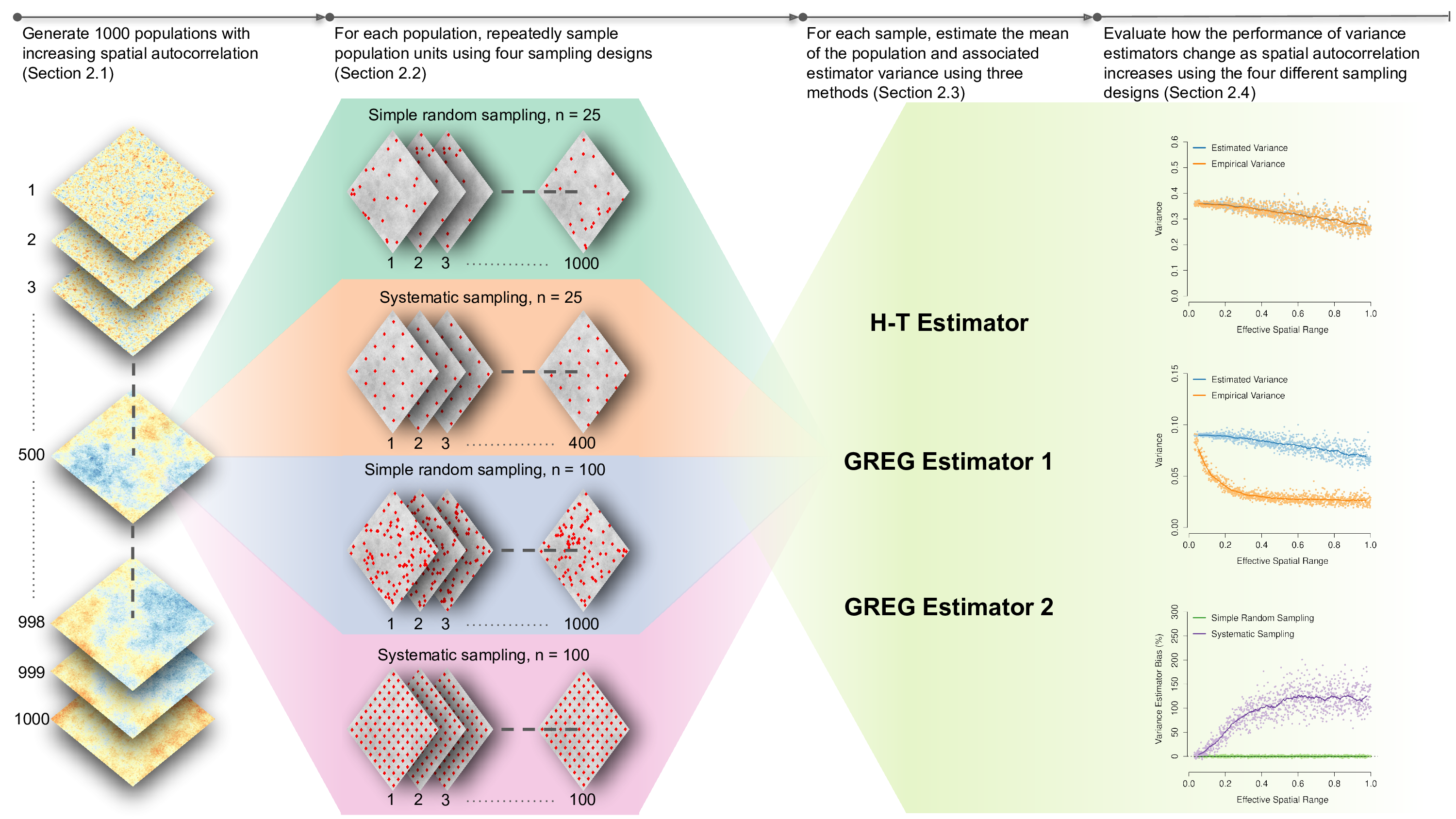}
\caption{Schematic detailing the synthetic populations analysis methods.}\label{schematic}
\end{figure}

The diagram in Figure \ref{schematic} lays out the methodology used to explore the effects of spatial autocorrelation on design-based inference following systematic and simple random sampling. Figure \ref{schematic} references subsequent sections where further analysis details are provided. In short, we generated a sequence of 1000 synthetic populations each exhibiting a different level of spatial autocorrelation. This was accomplished by setting up a sampling frame and a general super population model incorporating two spatially structured covariates. Then a sequence of 1000 synthetic populations of increasing spatial structure was created by lengthening the effective spatial range of the covariates for each successive population. For every synthetic population in the series, repeated sampling was performed following simple random and systematic sampling for two different sample sizes (four sampling designs in total). For every sample of population units, three population mean and variance estimator pairs were employed; each of which being at least approximately design-unbiased following simple random sampling given sufficient sample size: 1) An H-T estimator (\emph{H-T estimator}); 2) a GREG estimator using one covariate (\emph{GREG estimator 1}); 3) a GREG estimator using both covariates (\emph{GREG estimator 2}). The empirical variance of the estimators of the mean (i.e., variance of all possible sample means given a particular sampling design) was approximated by taking the variance of the estimated means of the repeated samples for each population.

\subsection{Synthetic populations}
A two-dimensional sampling frame was created by dividing a unit square into $N = 10\,000$ square grid cells of uniform size. This $100 \times 100$ grid serves as the enumeration of all possible sampling units for the synthetic populations. 

To generate synthetic populations, we first define a general super population model as
\begin{align}
\by &\sim \N(\beta_0 + \beta_1\bx_1 + \beta_2\bx_2, \tau^2\bI)\label{super-pop}\\
\bx_1 &\sim \N(\bzero, \bSigma)\label{super-pop-covar-1}\\
\bx_2 &\sim \N(\bzero, \bSigma)\label{super-pop-covar-2},
\end{align}
where $\by$ is an $N\times 1$ vector holding the values of the attribute of interest for every grid cell in the sampling frame. Equation \ref{super-pop} stipulates that $\by$ follows a multivariate normal distribution ($\N$) with an $N\times 1$ mean vector equal to the linear term, $\beta_0 + \beta_1\bx_1 + \beta_2\bx_2$. It can be seen that the linear term is dependent on two $N\times 1$ covariate vectors ($\bx_1$ and $\bx_2$) and three regression parameters ($\beta_0$, $\beta_1$ and $\beta_2$). The covariance matrix of the conditional distribution of $\by$ ($\tau^2\bI$) is $N \times N$ with off-diagonal elements equal to zero and diagonal elements equal to $\tau^2$. This covariance structure for $\by$ stipulates that the individual elements of $\by$ are independent after accounting for the linear term, $\beta_0 + \beta_1\bx_1 + \beta_2\bx_2$.

Sub-models \ref{super-pop-covar-1} and \ref{super-pop-covar-2} stipulate that the covariate vectors $\bx_1$ and $\bx_2$ follow a multivariate normal distribution with a zero mean ($\bzero$) and a covariance matrix $\bSigma$. To provide $\bx_1$ and $\bx_2$ (and subsequently $\by$) with spatial structure, the elements of $\bSigma$ were set using an exponential covariance function dependent on Euclidean separation distance between grid cell centers, i.e., $\bSigma_{ij} = \sigma^2\exp(-\phi \times d_{ij})$ where $d_{ij}$ is the distance between grid cell centers $i$ and $j$. The parameters $\sigma^2$ and $\phi$ control the spatial variance and decay of the process, respectively. We define \emph{effective spatial range} ($esr$) to be the separation distance at which the spatial correlation between locations drops to 0.05, i.e., $esr = -\ln(0.05)/\phi \approx 3/\phi$.

To generate a synthetic population from the general super population model shown in Equation \ref{super-pop}, values need to be assigned to all parameters ($\beta_0, \beta_1, \beta_2, \tau^2, \sigma^2 \text{ and } \phi$), then a random number generator can be used to draw a sample from the super population distribution. This sample then serves as a synthetic population. 

\begin{figure}[!h]
\centering
\includegraphics[width = .75\textwidth]{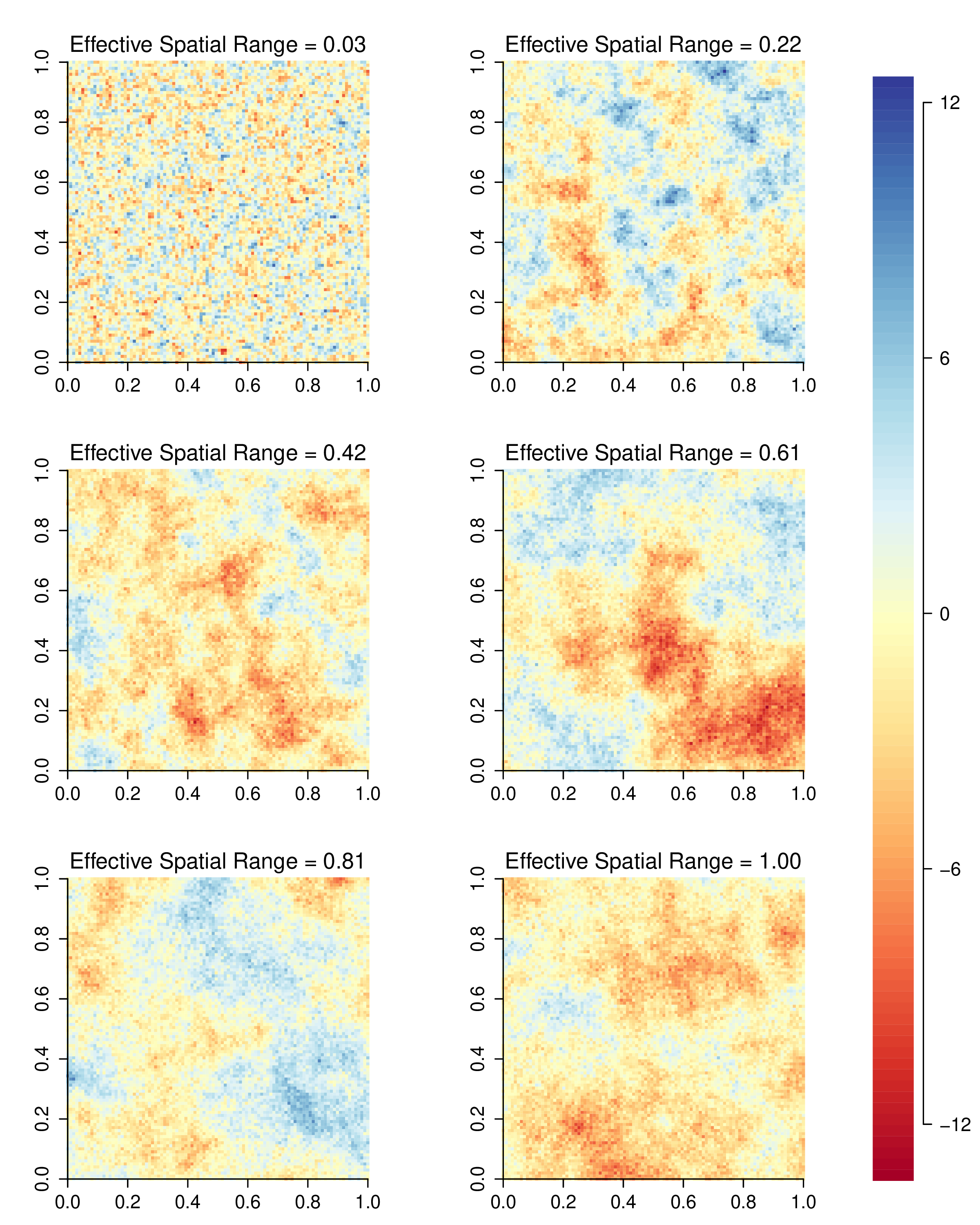}
\caption{Maps of six synthetic populations showing the change in spatial structure as effective spatial range increases.}\label{pops}
\end{figure}

We generated a series of 1000 synthetic populations with successive populations exhibiting increased spatial autocorrelation. To accomplish this, we set $\beta_0, \beta_1, \beta_2, \tau^2$ and $\sigma^2$ equal to $0,1,1,1$ and $4$, respectively, and held these parameters constant across the 1000 simulated populations. For the first population draw, $esr$ was set to $\approx 0.03$, i.e., $\phi$ was set to $3/0.03 = 100$. To increase spatial autocorrelation, we incrementally raised $esr$ by $\approx 0.00097$ for each successive synthetic population draw. This process resulted in the generation of a series of 1000 populations where $esr$ of the covariates $\bx_1$ and $\bx_2$ increased from $0.03$ to $1$ uniformly on the unit square. Figure \ref{pops} displays six of the $\by$ synthetic populations in the series, showing that, in general, the strength of the spatial structure of $\by$ increases as $esr$ increases. The authors note the general super population model employed here has two parameters controlling spatial autocorrelation, $\sigma^2$ and $\phi$, but for this analysis only the effects of changing $\phi$ are examined. A future simulation study could examine the influence of altering either $\sigma^2$ alone or $\sigma^2$ and $\phi$ together to further assess how changes in spatial autocorrelation effect design-based inference following systematic sampling. 

\subsection{Sampling designs}
Simple random and systematic repeated sampling was used to generate empirical sampling distributions of three candidate estimators for each synthetic population. The three candidate estimators are described in detail in Section \ref{candidates}. For the simple random sampling designs, 1000 sample sets of size $n$ were randomly selected without replacement. For the systematic sampling designs, sample sets were selected in such a way that the selected sample units formed a $\sqrt{n} \times \sqrt{n}$ regular grid of cells. This is an equal probability sampling design and allows for the selection of all possible samples of size $n$. The total number of possible sample sets is $N/n$ for this type of systematic design. For this analysis, we examined sample sizes $n = 25$ and $100$.

\subsection{Candidate estimators}\label{candidates}
The general framework used to derive the three specific estimators for $\mu$ (the true population mean) examined in this study is 
\begin{equation}
 \hat{\mu} = \frac{\bones_N^\top\hat{\by}_U}{N} + \frac{\bones_n^\top(\hat{\by}_s-\by_s)}{n}\label{ybar}
\end{equation}
where $\by_s$ is an $n \times 1$ vector of observed values and $\hat{\by}_s$ contains associated model fitted values. $\by_U$ is an $N \times 1$ vector holding model predictions for all grid cells in the population. $\bones_n$ and $\bones_N$ are $n \times 1$ and $N \times 1$ vectors of ones, respectively. The superscript $^\top$ denotes the transpose operator. Stated simply, Equation \ref{ybar} specifies that $\hat{\mu}$ equals the mean of model predictions for all grid cells plus the mean of model residuals. We note that the assisting models used for the GREG estimators in the synthetic populations analysis were fitted in a way that ensures that the sum of the residuals equals zero, making the second term in Equation \ref{ybar} unnecessary. However, in the Harvard Forest stem map analysis described in Section \ref{hf-analysis}, $\by$ is squareroot transformed before modeling fitting, requiring us to employ the second term in Equation \ref{ybar} to remain approximately design-unbiased. The general variance estimator employed here is 
\begin{equation}
\widehat{Var}(\hat{\mu}) = s^2/n\label{yse}
\end{equation}
where
\begin{equation}
s^2 = \frac{\bones_n^\top(\hat{\by}_s-\by_s)^2}{n-p}.\label{ys2}
\end{equation}

In Equation \ref{ys2}, $p$ equals the number of parameters to be estimated in the assisting model. Ignoring the sampling fraction, $n/N$, Equation \ref{yse} unbiasedly estimates $Var(\hat{\mu})$ following simple random sampling. We derived three candidate estimators for the population mean and its variance from Equations \ref{ybar} and \ref{yse}, respectively, by specifying three different assisting models to obtain $\hat{\by}_s$ and $\hat{\by}_U$. The \emph{H-T estimator} was defined by setting
\begin{align}
\hat{\by}_s & = \hat{\beta}_0\bones_n\qquad \text{and} \label{db-fit} \\
\hat{\by}_U & = \hat{\beta}_0\bones_N. \label{db-pred} 
\end{align}
The \emph{GREG estimator 1} was defined using
\begin{align}
\hat{\by}_s & = \hat{\beta}_0\bones_n + \hat{\beta}_1\bx_{1s}\qquad \text{and} \label{ma1-fit}\\
\hat{\by}_U & = \hat{\beta}_0\bones_N + \hat{\beta}_1\bx_{1}. \label{ma1-pred}
\end{align}
The \emph{GREG estimator 2} was defined using
\begin{align}
\hat{\by}_s & = \hat{\beta}_0\bones_n + \hat{\beta}_1\bx_{1s} + \hat{\beta}_2\bx_{2s}\qquad \text{and} \label{ma2-fit}\\
\hat{\by}_U & = \hat{\beta}_0\bones_N + \hat{\beta}_1\bx_{1} + \hat{\beta}_2\bx_{2}. \label{ma2-pred}
\end{align}
The $\hat{\beta}$'s are ordinary least squares estimates based on the sample. The $n \times 1$ vectors $\bx_{1s}$ and $\bx_{2s}$ are covariate values for the sample locations. In the Appendix, we demonstrate that when using the assisting model shown in Equations \ref{db-fit} and \ref{db-pred}, Equations \ref{ybar} and \ref{yse} become the typical H-T estimators for the population mean and its variance, respectively.

\subsection{Summary statistics}
Several summary metrics were calculated for each synthetic population, candidate estimator and sampling design combination to aid in the assessment of the simulation study results. Empirical variances were obtained by calculating the variance of the means estimated for the samples from each simulated population. The mean of the estimated variances was also calculated. Subsequent figures compare these two metrics to assess how well the variance estimators approximate the empirical variance. We
also calculated the relative bias of the variance estimators.

\section{Harvard Forest stem map analysis}\label{hf-analysis}
\subsection{Dataset description}\label{hf-data}
\begin{figure}[!h]
\centering
\captionsetup[subfigure]{justification=centering}
\subfloat[Tree locations colored by aboveground biomass (AGB).]{\includegraphics[width=.5\textwidth]{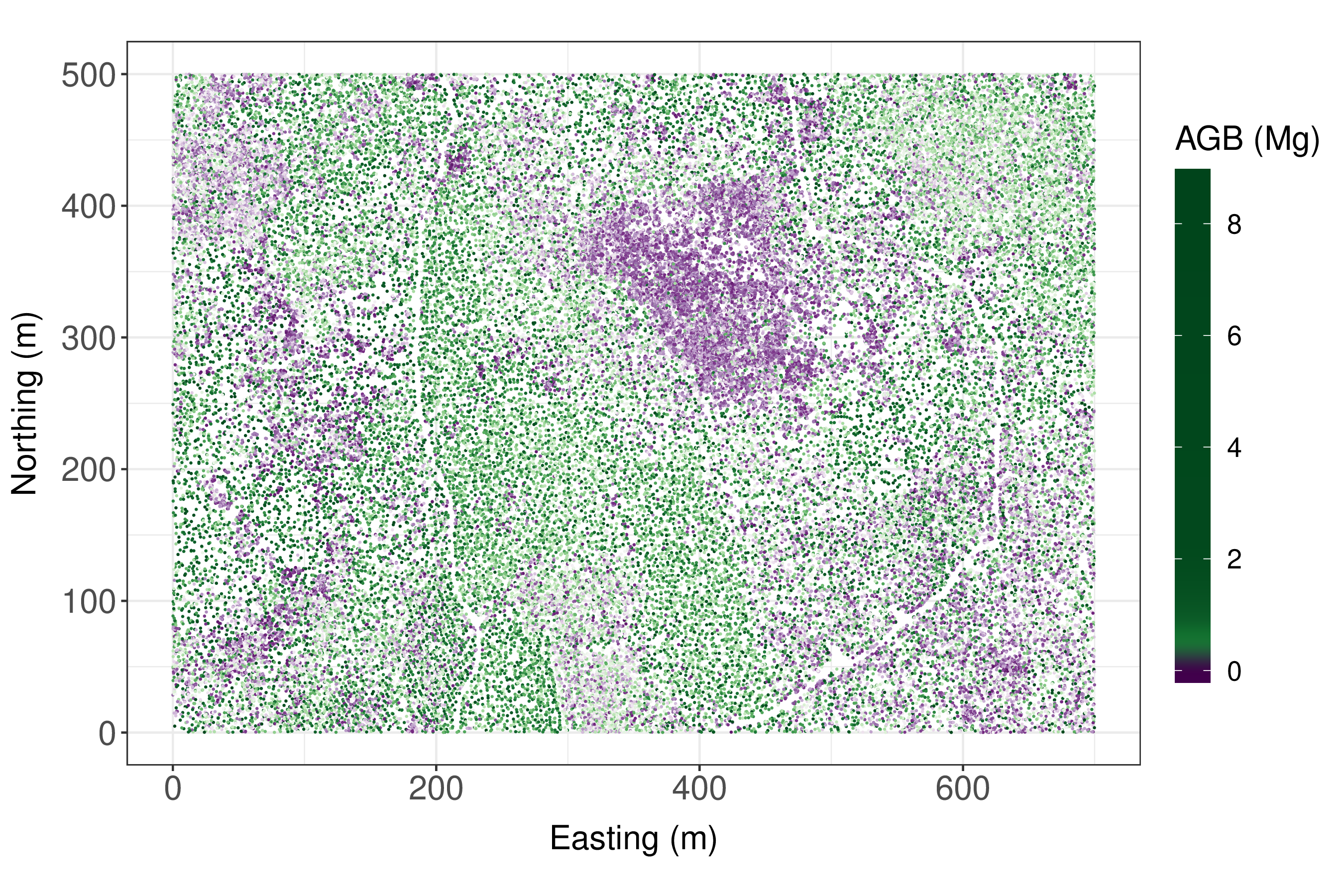}}
\subfloat[Lidar-derived 90th percentile height (P90).]{\includegraphics[width=.5\textwidth]{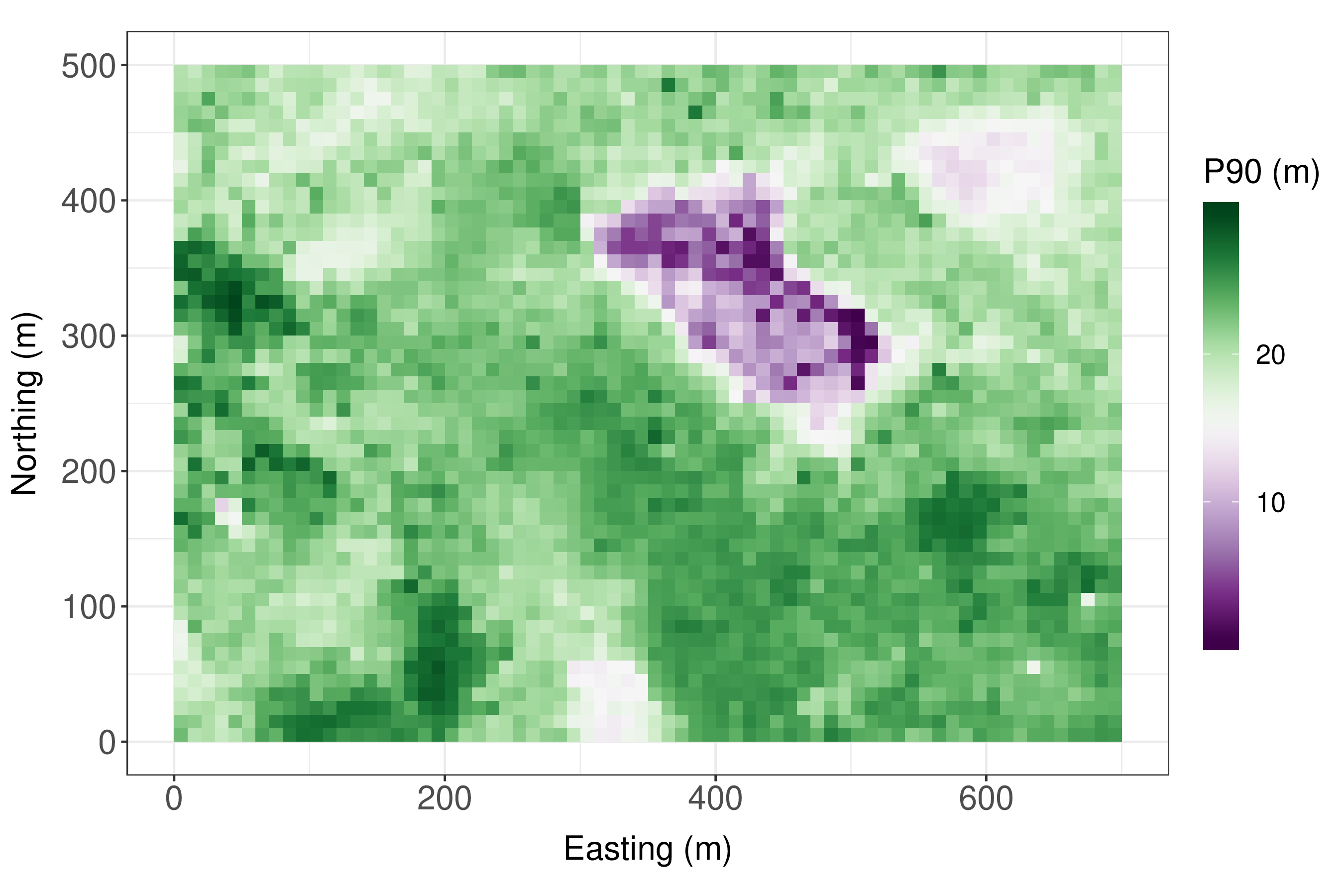}}\\
\subfloat[Lidar-derived 10th percentile height (P10).]{\includegraphics[width=.5\textwidth]{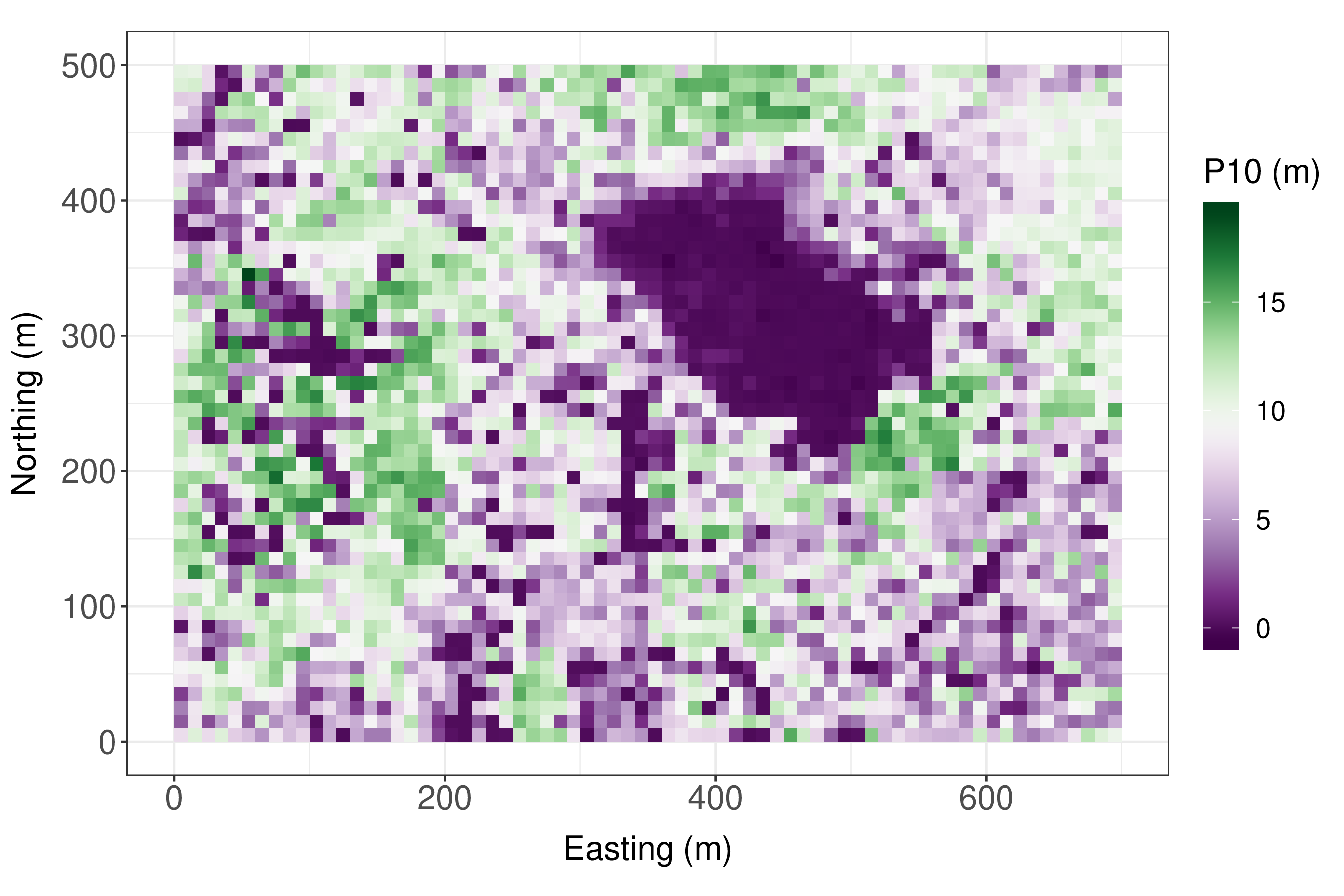}}
\subfloat[Hyperspectral-derived normalized vegetation index (NDVI).]{\includegraphics[width=.5\textwidth]{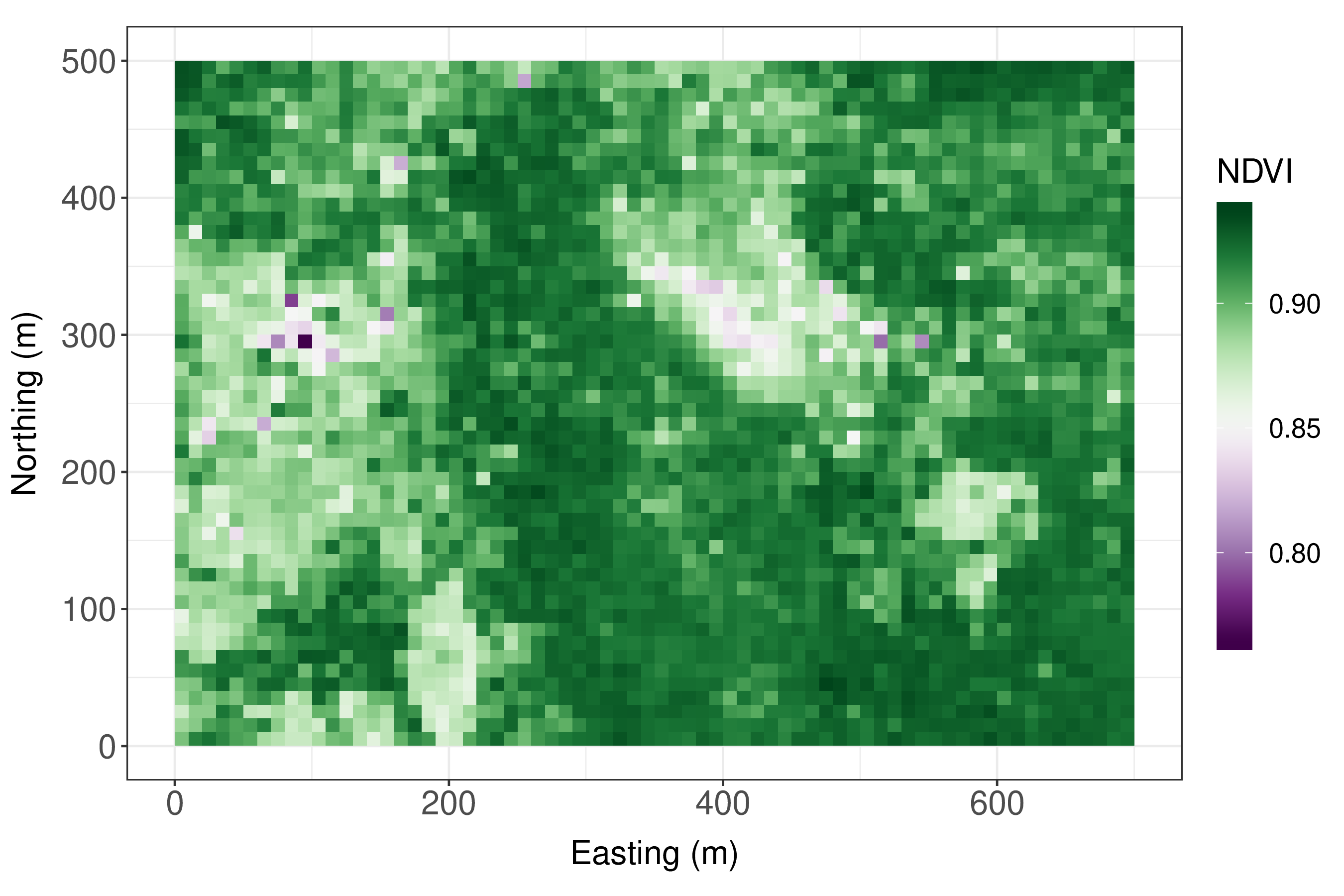}}
\caption{Harvard Forest tree and covariate maps.}\label{hf-maps}
\end{figure}

A complete census of all woody stems $\ge 1$ cm in diameter at breast height was conducted on a 35-ha plot between June of 2010 and March of 2014 within the Prospect Hill tract of Harvard Forest in Massachusetts. Figure \ref{hf-maps}a marks tree locations within the study area colored according to aboveground biomass (AGB). A total of $83\,801$ stem-mapped trees were used in the analysis. NASA Goddard's Lidar, Hyperspectral and Thermal (G-LiHT) imager was flown over the stem map area in June of 2012 \citep{cook2013}. The lidar and hyperspectral data obtained was summarized to a 10m x 10m resolution ($N = 3500$) and a set of $\approx 70$ potential covariates were derived. The remote sensing covariate set comprised lidar-based percentile heights and relative densities in addition to several hyperspectral-based vegetation indicies including normalized vegetation index, red edge and others. 

\subsection{Sampling designs}
Similar to the analysis of synthetic populations described in Section \ref{synth-description}, simple random and systematic repeated sampling was used to generate empirical sampling distributions for the Harvard Forest AGB density (Mg/ha) candidate estimators (these estimators are described fully in Section \ref{hf-candidates}). To avoid confounding factors effecting estimators related to small population size, we adopted an infinite population sampling frame where circular plot centers were selected using simple random (systematic) sampling. The plot radius was set to $\approx 5.64$ $(\sqrt{100/\pi})$ so the plot area matched the remote sensing grid cell areas (100m$^2$). We note that the simple random and systematic sampling of plots was done in such a way as to ensure that no plots within a sample overlapped each other or the stem map area boundary. For each selected circular plot, AGB density (Mg/ha) was calculated by summing the AGB for all measured trees within the plot and dividing by the plot area. The same lidar and hyperspectral covariates calculated for the gridded dataset (described in Section \ref{hf-data}) were calculated over each plot. 10\,000 samples of size $n$ were selected with each sampling design. Sample sizes $n = 35 (5\times7)$ and $140 (10\times140)$ were examined for the Harvard Forest analysis. Sample sizes examined for the Harvard Forest analysis were different from the synthetic populations analysis because the Harvard Forest domain is rectangular and we wanted to ensure consistent spatial coverage for the systematic samples.

\subsection{Candidate estimators}\label{hf-candidates}
\begin{figure}[!h]
\centering
\includegraphics[width = \textwidth]{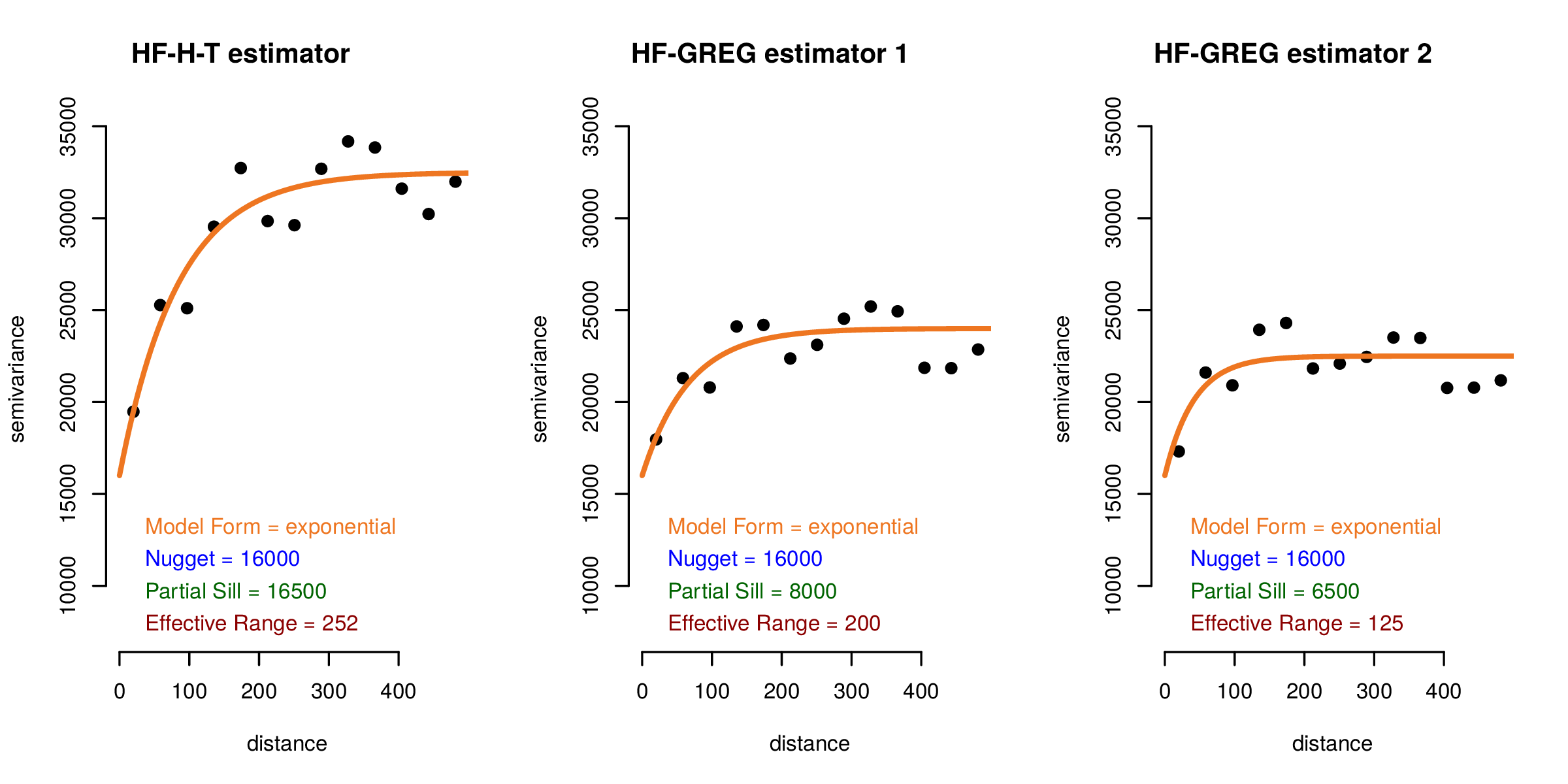}
\caption{Semivariograms and semivariogram models fitted to residuals of Harvard Forest candidate estimators for one simple random sampling set (n = 140). These plots demonstrate the tendency that residual spatial autocorrelation decreases as we move to the more complex model (\emph{HF-GREG estimator 2}).}\label{varios}
\end{figure}

We derived three candidate estimators for the Harvard Forest simulated sampling analysis using the same general framework described in Equations \ref{ybar} and \ref{yse} by specifying three different assisting models to obtain $\hat{\by}_s$ and $\hat{\by}_U$. The \emph{HF-H-T estimator} was defined by setting
\begin{align}
\hat{\by}_s & = \hat{\beta}_0\bones_n\qquad \text{and} \label{hf-db-fit} \\
\hat{\by}_U & = \hat{\beta}_0\bones_N.\label{hf-db-pred}
\end{align}
The \emph{HF-GREG estimator 1} was defined using
\begin{align}
\hat{\by}_s & = (\hat{\beta}_0\bones_n + \hat{\beta}_1\mathbf{P90}_{s})^2\qquad \text{and} \label{hf-ma1-fit} \\
\hat{\by}_U & = (\hat{\beta}_0\bones_N + \hat{\beta}_1\mathbf{P90})^2.\label{hf-ma1-pred}
\end{align}
The \emph{HF-GREG estimator 2} was defined using
\begin{align}
\hat{\by}_s & = (\hat{\beta}_0\bones_n + \hat{\beta}_1\mathbf{P90}_{s} + \hat{\beta}_2\mathbf{P10}_{s} + \hat{\beta}_3\mathbf{NDVI}_{s})^2 \qquad \text{and} \label{hf-ma2-fit} \\
\hat{\by}_U & = (\hat{\beta}_0\bones_N + \hat{\beta}_1\mathbf{P90} + \hat{\beta}_2\mathbf{P10} + \hat{\beta}_3\mathbf{NDVI})^2.\label{hf-ma2-pred}
\end{align}
The $\hat{\beta}$'s are ordinary least squares estimates based on the sample (a squareroot transformation was used to improve model fits). The $N \times 1$ covariate vectors $\mathbf{P90}$, $\mathbf{P10}$ and $\mathbf{NDVI}$ hold remote sensing values for the 3500 grid cells. Specifically, $\mathbf{P90}$ and $\mathbf{P10}$ are lidar-derived 90th and 10th percentile heights, respectively. $\mathbf{NDVI}$ is hyperspectral-derived normalized vegetation index. Cloropleth maps of these three covariates are shown in Figure \ref{hf-maps}. $\mathbf{P90}_s$, $\mathbf{P10}_s$ and $\mathbf{NDVI}_s$ are corresponding remote sensing covariates over selected plot locations. Covariates for models were selected based on preliminary model fitting exercises where we sought to build two models with good fit statistics and low residual spatial autocorrelation. Figure \ref{varios} provides evidence that \emph{HF-GREG estimator 1} and \emph{HF-GREG estimator 2} has weaker residual spatial autocorrelation than \emph{HF-H-T estimator}, i.e., \emph{esr} tends to be shorter and $\sigma^2$ tends to be lower in the GREG estimator residuals. Further, \emph{HF-GREG estimator 2} shows less residual spatial structure than \emph{HF-GREG estimator 1}. 

\subsection{Summary statistics}
Similar summary statistics used for the synthetic populations analysis were calculated for the Harvard Forest candidate estimator and sampling design combinations, including empirical variances, means of estimated variances and variance estimator bias. These results are presented in a subsequent results table.

\section{Results and Discussion}
\subsection{Synthetic populations analysis}
\begin{figure}[!h]
\centering
\subfloat[][\parbox{.2\textwidth}{H-T estimator \\ sample size = 25}]{\includegraphics[width=.33\textwidth]{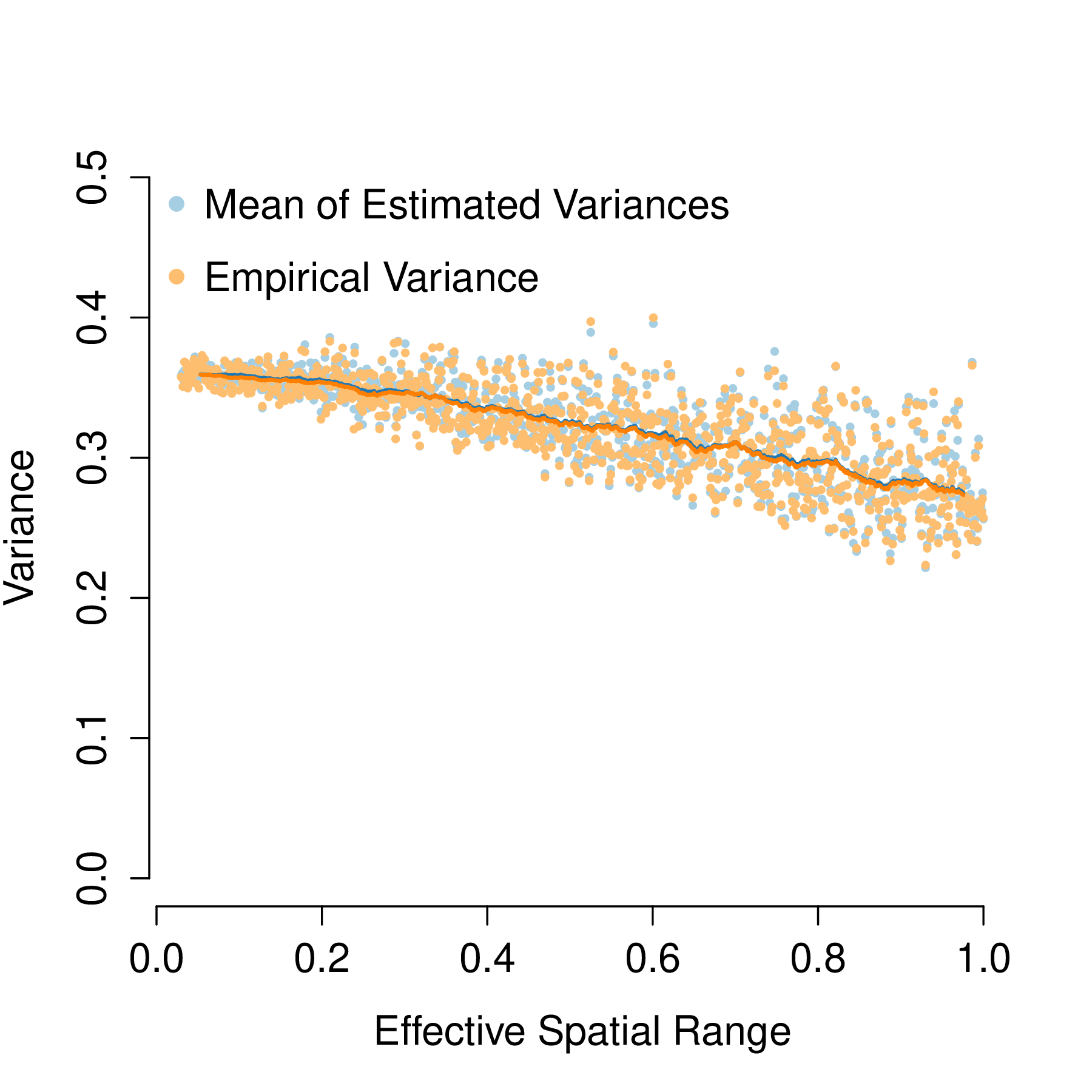}}
\subfloat[][\parbox{.2\textwidth}{GREG estimator 1 \\ sample size = 25}]{\includegraphics[width=.33\textwidth]{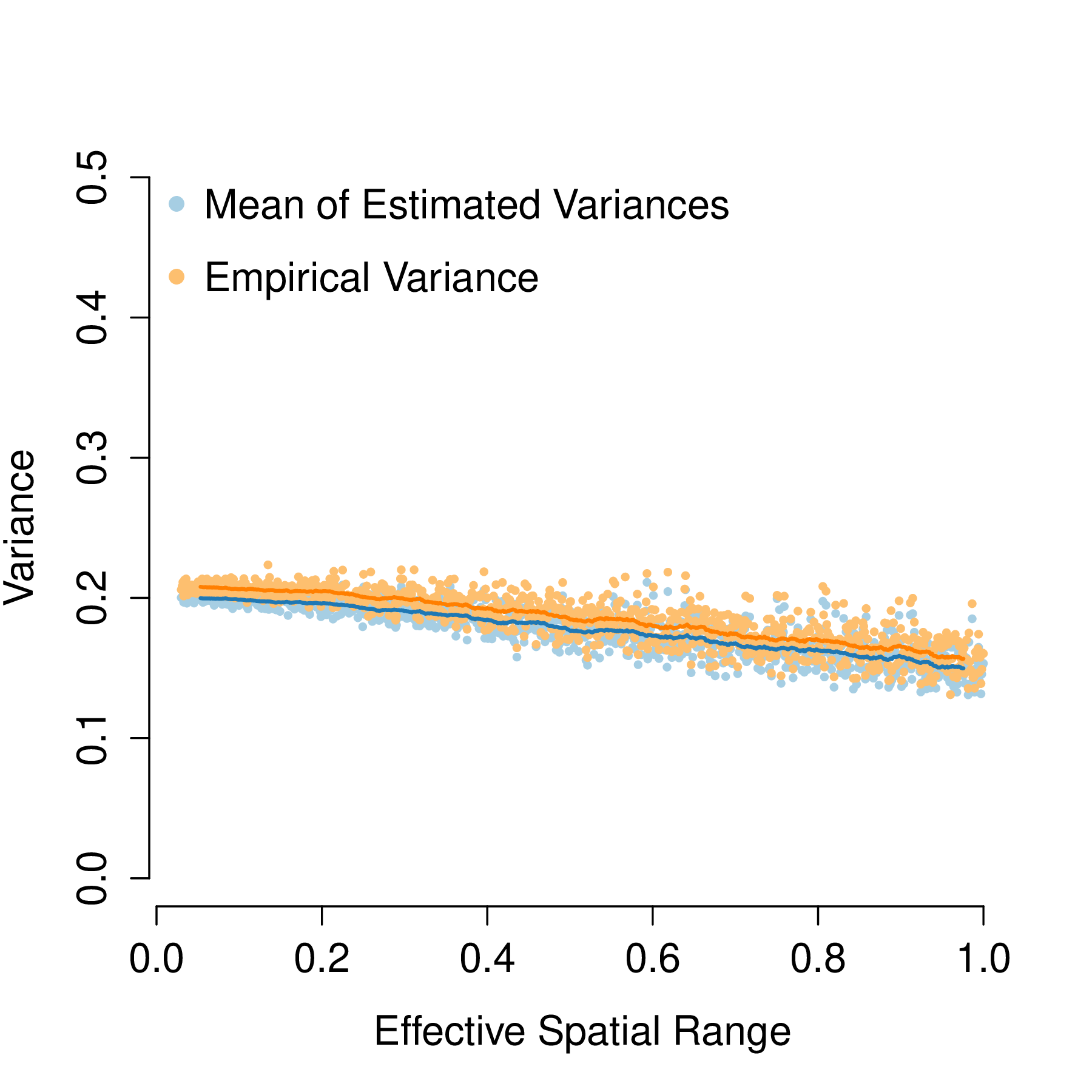}}
\subfloat[][\parbox{.2\textwidth}{GREG estimator 2 \\ sample size = 25}]{\includegraphics[width=.33\textwidth]{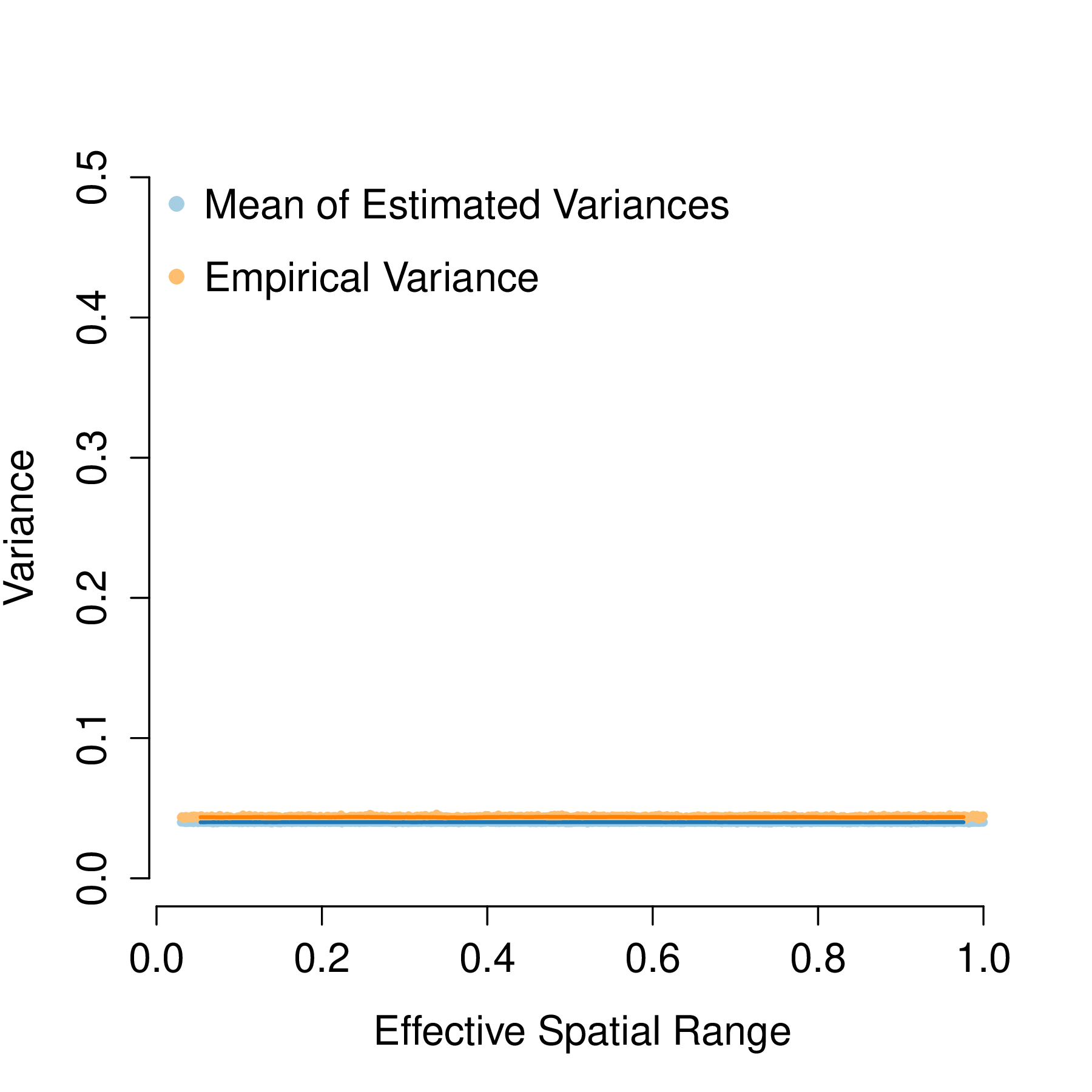}}\\
\subfloat[][\parbox{.2\textwidth}{H-T estimator \\ sample size = 100}]{\includegraphics[width=.33\textwidth]{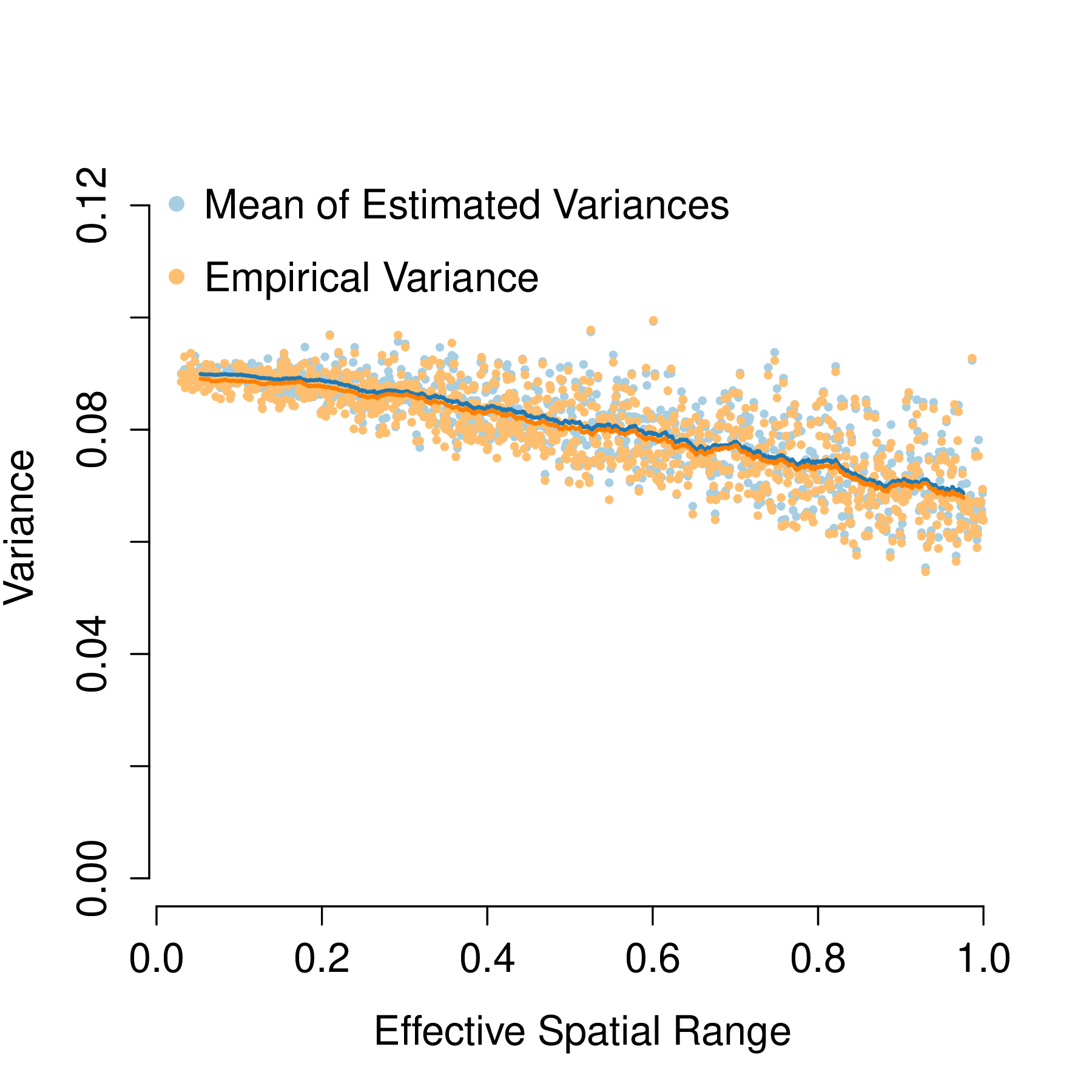}}
\subfloat[][\parbox{.2\textwidth}{GREG estimator 1 \\ sample size = 100}]{\includegraphics[width=.33\textwidth]{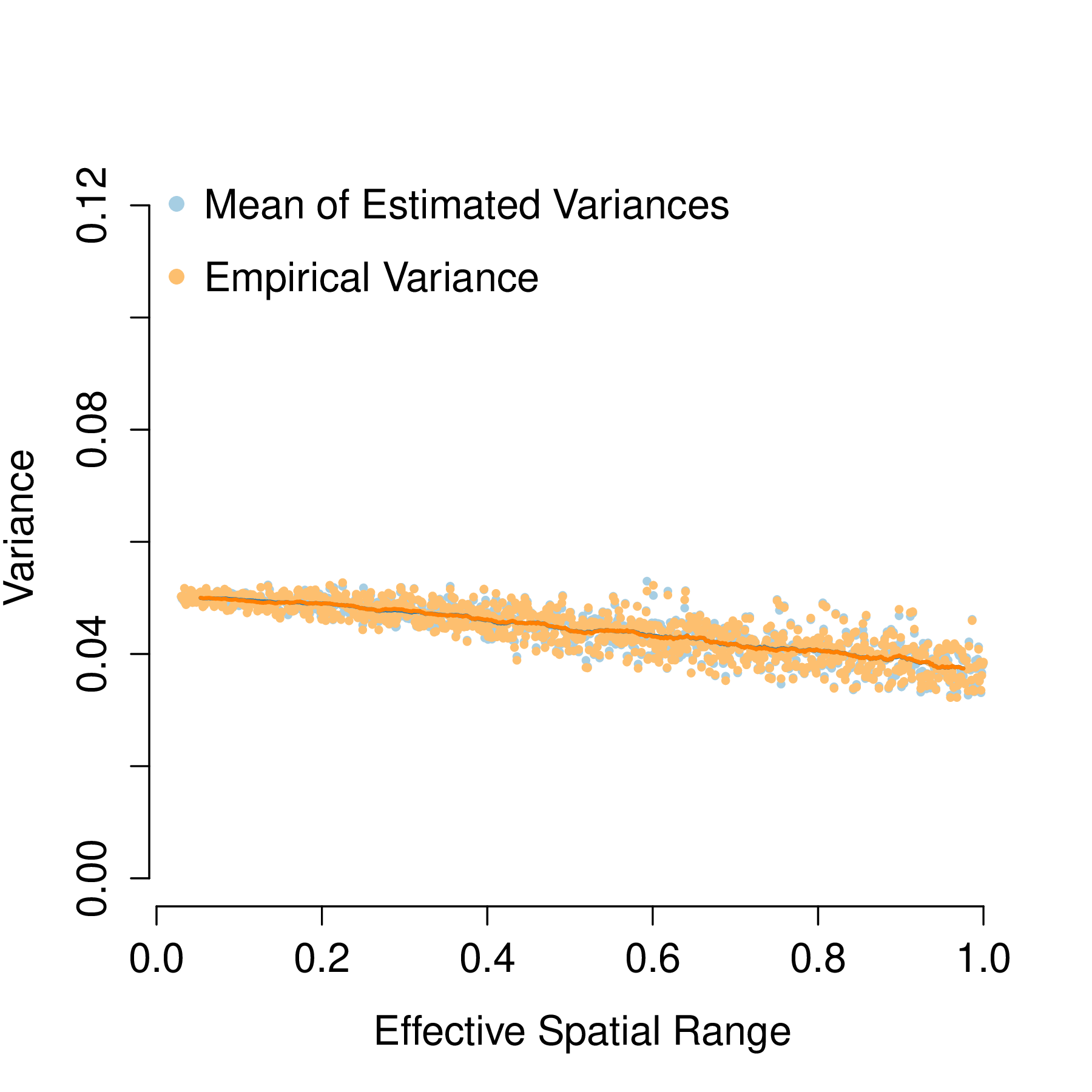}}
\subfloat[][\parbox{.2\textwidth}{GREG estimator 2 \\ sample size = 100}]{\includegraphics[width=.33\textwidth]{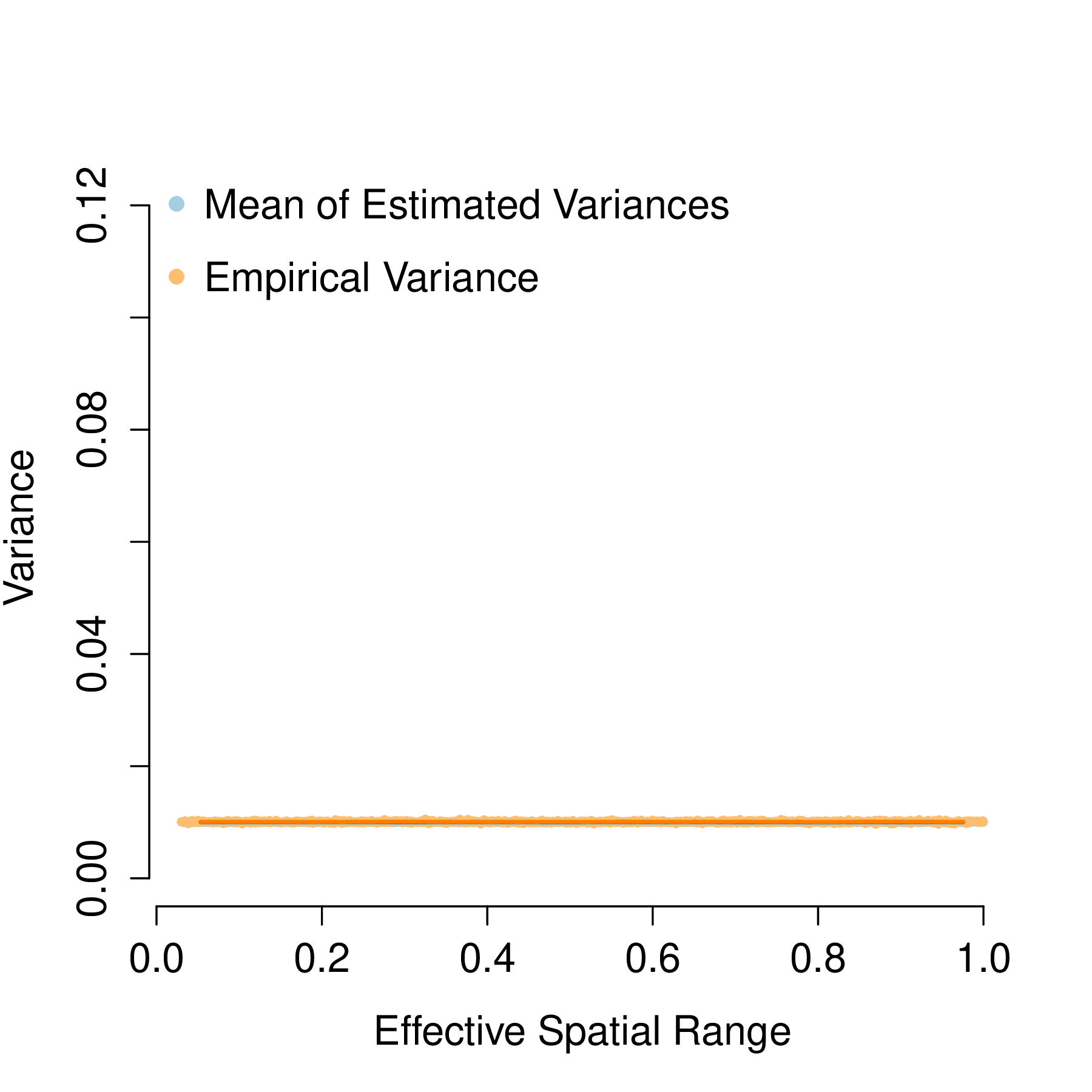}}
\caption{Comparisons between empirical and mean of estimated variances following simple random sampling. Orange and blue lines delineate the moving average of empirical and mean of estimated variances, respectively.}\label{srs-synthetic-results}
\end{figure}

\begin{figure}[!h]
\centering
\subfloat[][\parbox{.2\textwidth}{H-T estimator \\ sample size = 25}]{\includegraphics[width=.33\textwidth]{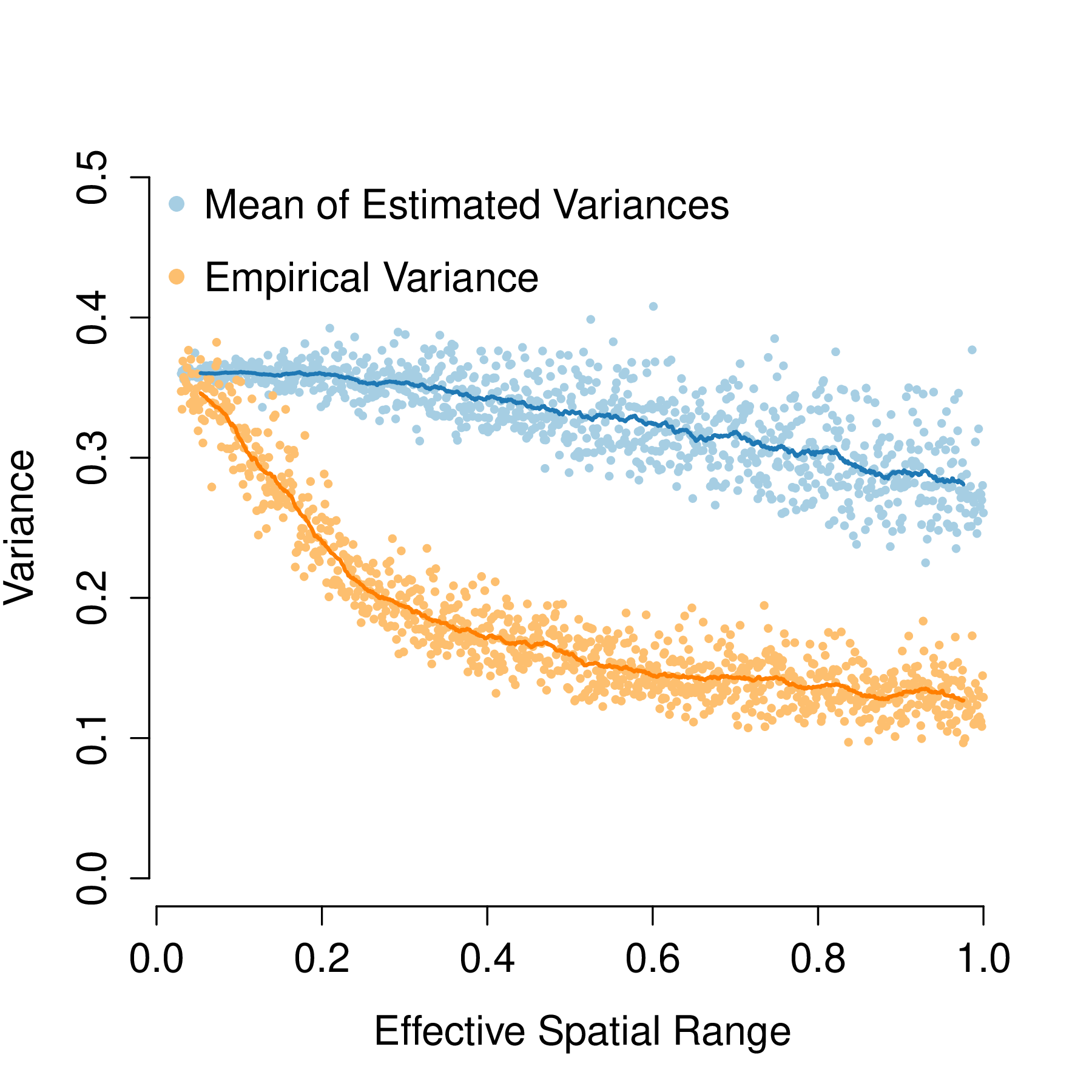}}
\subfloat[][\parbox{.2\textwidth}{GREG estimator 1 \\ sample size = 25}]{\includegraphics[width=.33\textwidth]{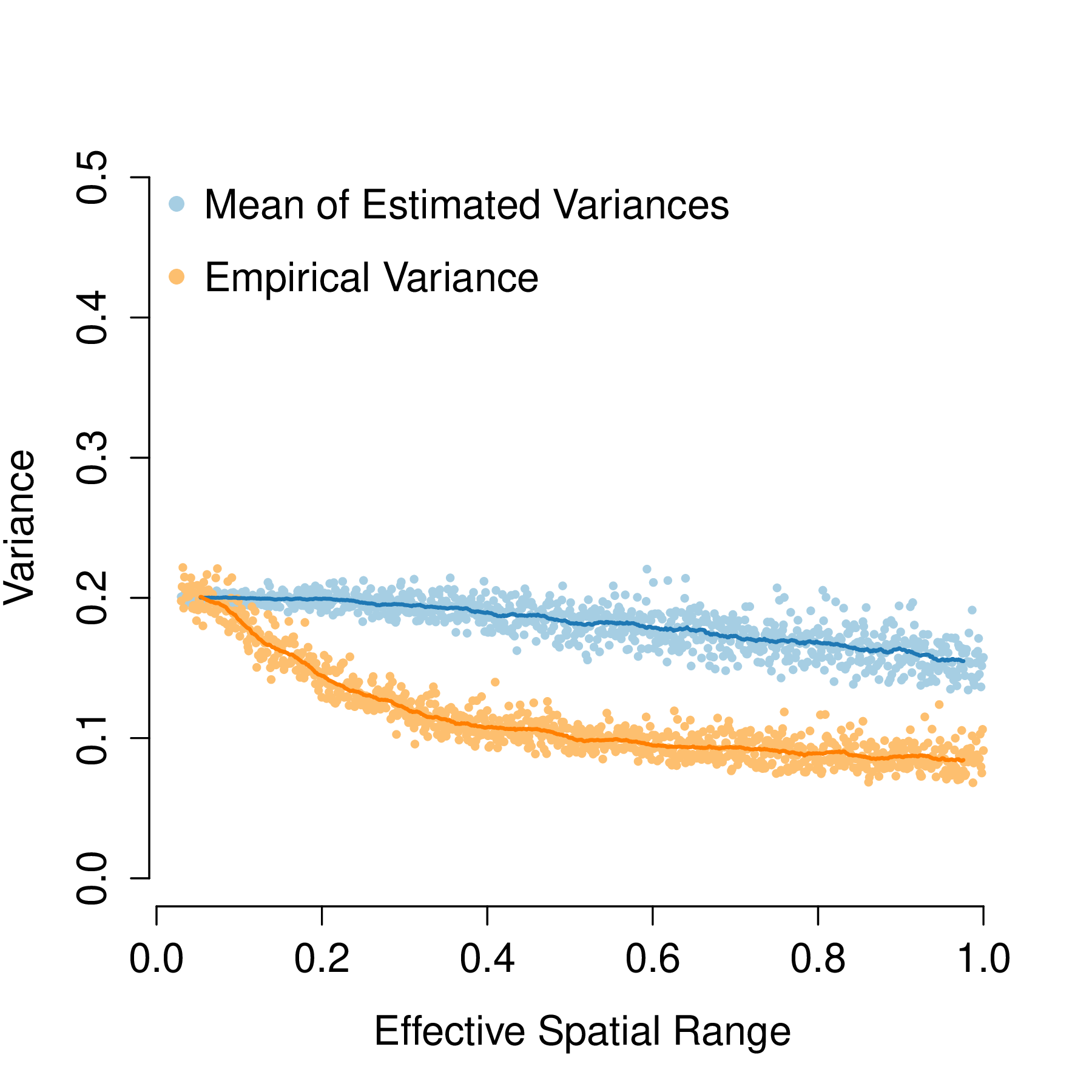}}
\subfloat[][\parbox{.2\textwidth}{GREG estimator 2 \\ sample size = 25}]{\includegraphics[width=.33\textwidth]{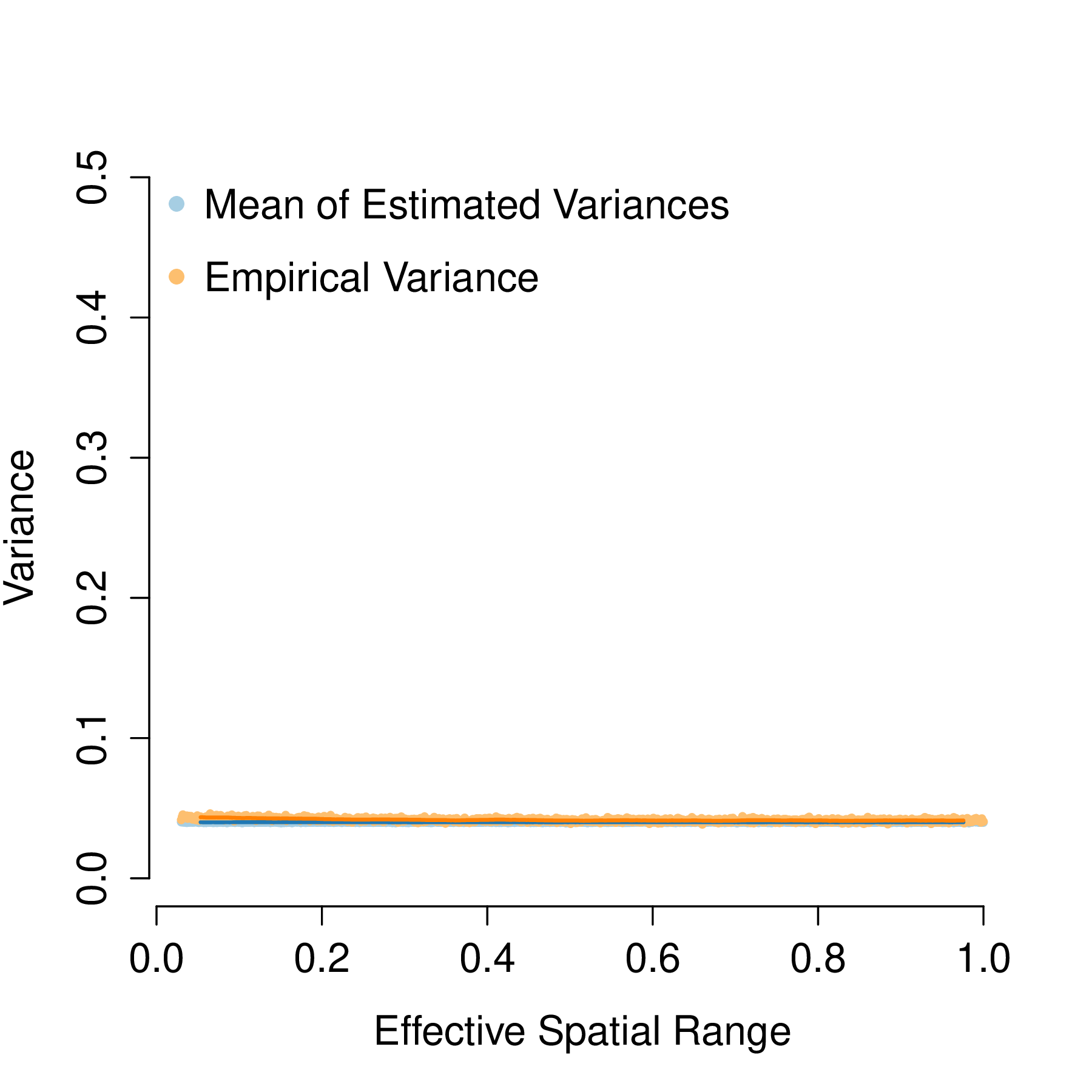}}\\
\subfloat[][\parbox{.2\textwidth}{H-T estimator \\ sample size = 100}]{\includegraphics[width=.33\textwidth]{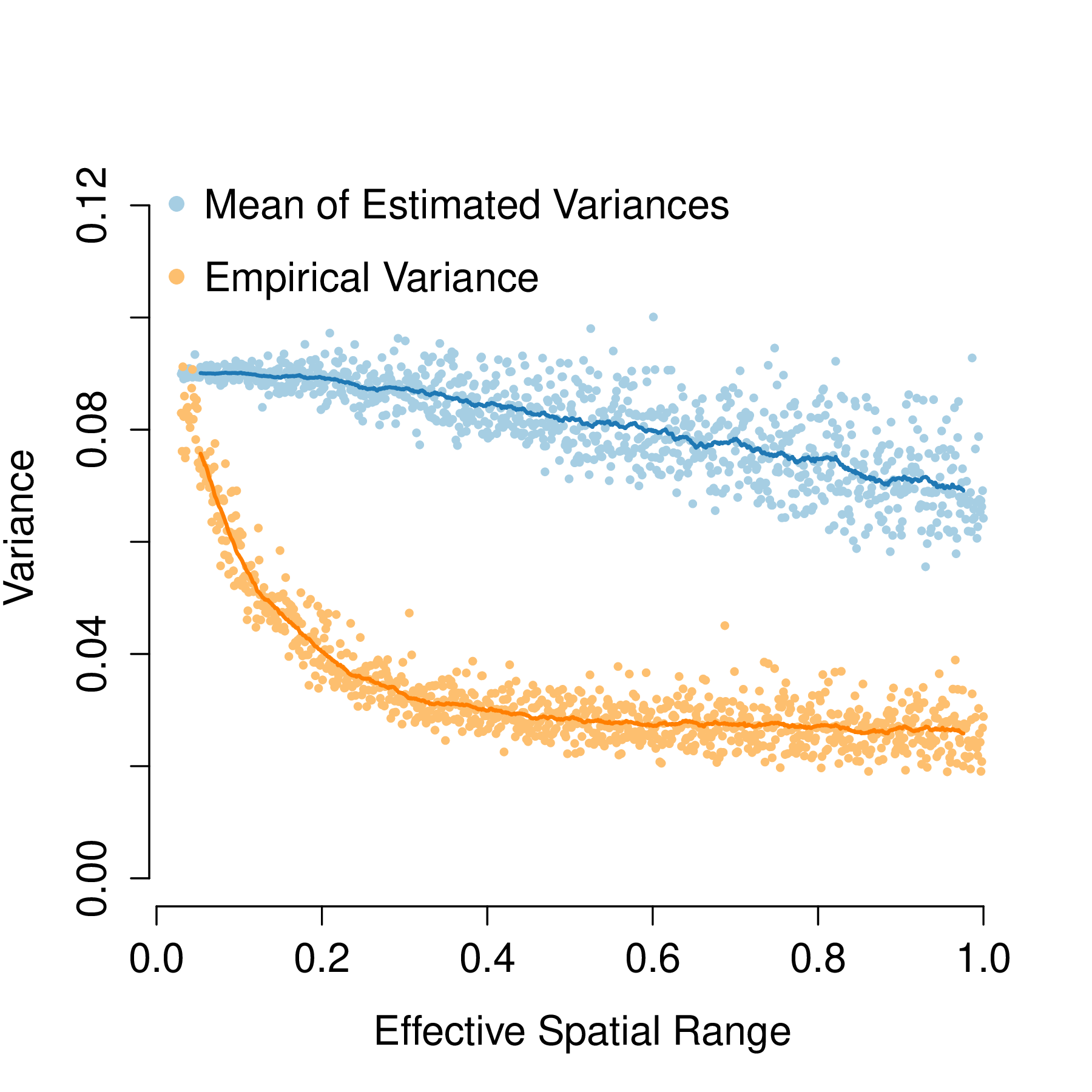}}
\subfloat[][\parbox{.2\textwidth}{GREG estimator 1 \\ sample size = 100}]{\includegraphics[width=.33\textwidth]{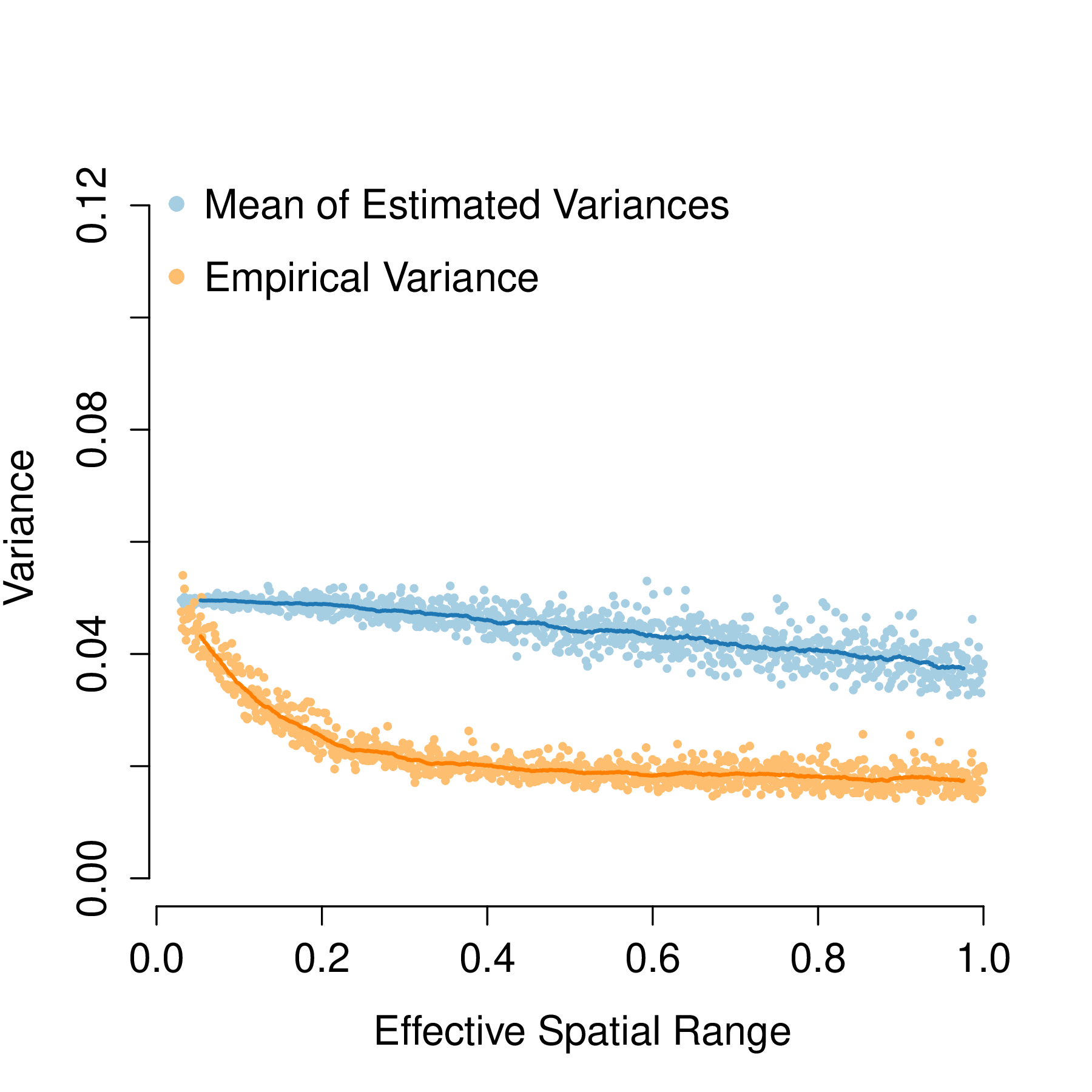}}
\subfloat[][\parbox{.2\textwidth}{GREG estimator 2 \\ sample size = 100}]{\includegraphics[width=.33\textwidth]{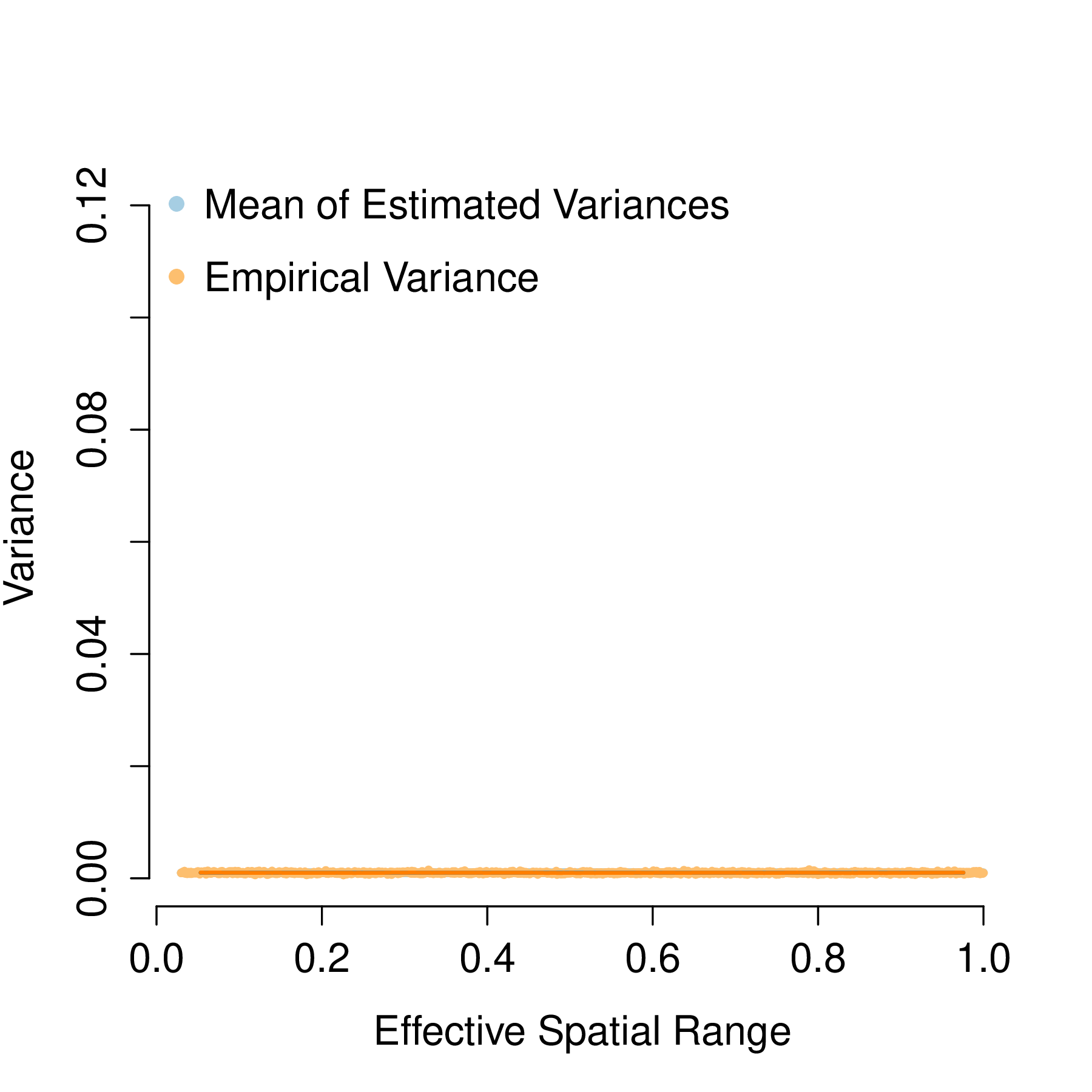}}
\caption{Comparisons between empirical and mean of estimated variances following systematic sampling. Orange and blue lines delineate the moving average of empirical and mean of estimated variances, respectively.}\label{sys-synthetic-results}
\end{figure}

Figure \ref{srs-synthetic-results} plots the mean of the estimated variances for each of the 1000 synthetic populations as blue points and the associated empirical variances as orange points for the simple random sampling designs. Figure \ref{srs-synthetic-results} shows close agreement between the empirical and mean of estimated variances for the three estimators across every level of spatial autocorrelation examined. This confirms that spatial autocorrelation does not affect the performance of the variance estimators examined here when simple random sampling is used to select sample observations. However, Figure \ref{sys-synthetic-results} (the complementary figure for the systematic sampling designs) shows a divergence between the empirical and mean of estimated variances for the \emph{H-T estimator}. The empirical variance for this estimator tends to drift lower than the mean of estimated variances as spatial autocorrelation increases (Figures \ref{sys-synthetic-results}a and d). This suggests that spatial autocorrelation in the population can cause variance estimators that are unbiased following simple random sampling to be conservative when used with systematic samples. This phenomenon is present for \emph{GREG estimator 1} as well, although, the separation between the empirical and mean of estimated variances is less substantial (Figures \ref{sys-synthetic-results}b and e). For \emph{GREG estimator 2}, the empirical and mean of estimated variances closely align as they did for the simple random sampling case shown in Figure \ref{srs-synthetic-results}. The variance estimator used for \emph{GREG estimator 1} and \emph{GREG estimator 2} relies on the variance of the assisting model residuals rather than the variance of the sample observations as with the \emph{H-T estimator}. Considering that residual spatial autocorrelation is reduced when $\bx_1$ is used and completely eliminated when both $\bx_1$ and $\bx_2$ are used in the assisting model, our results support the notion that reducing spatial autocorrelation in assisting model residuals reduces estimator variance over-estimation, following systematic sampling.

\begin{figure}[!h]
\centering
\subfloat[][\parbox{.2\textwidth}{H-T estimator \\ sample size = 25}]{\includegraphics[width=.33\textwidth]{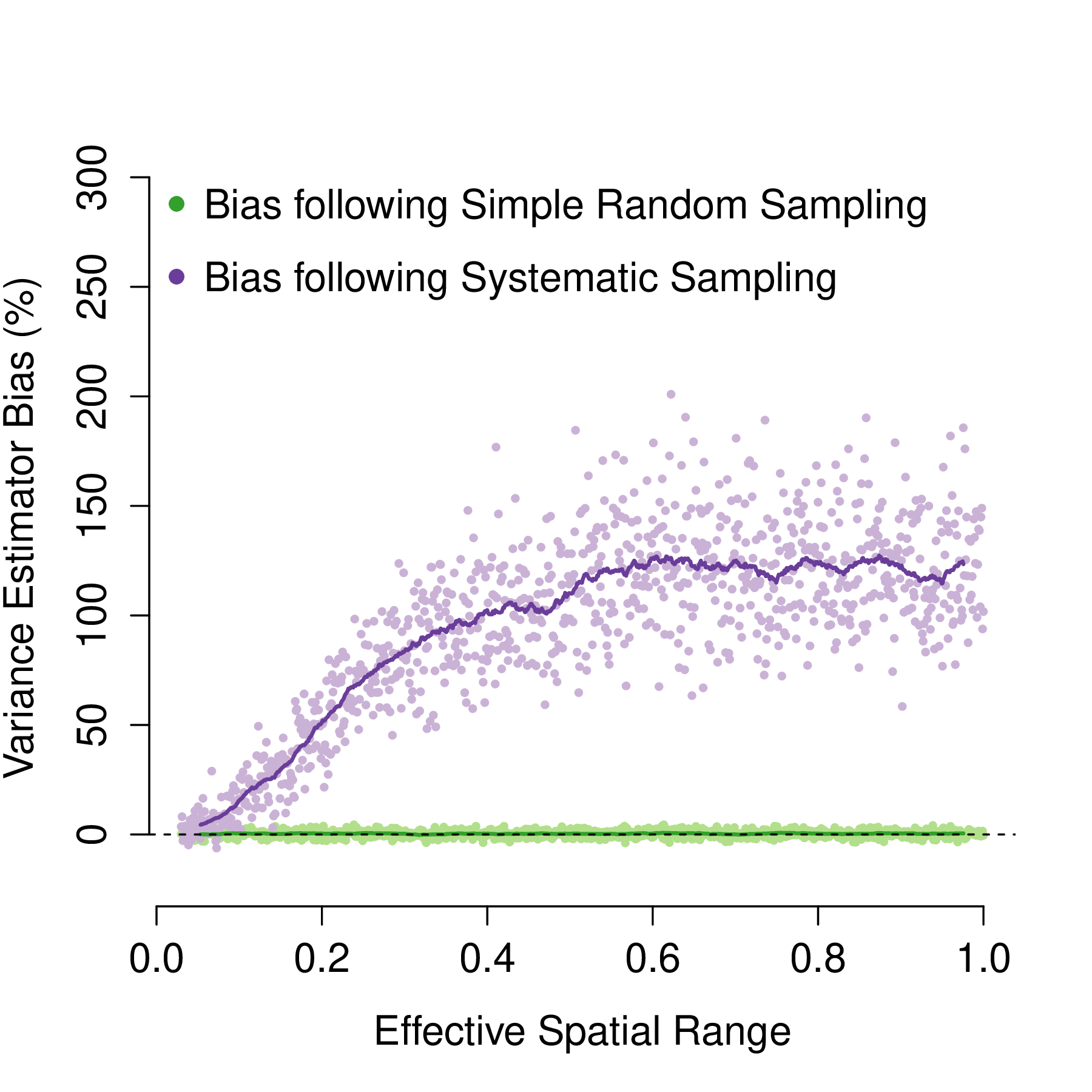}}
\subfloat[][\parbox{.2\textwidth}{GREG estimator 1 \\ sample size = 25}]{\includegraphics[width=.33\textwidth]{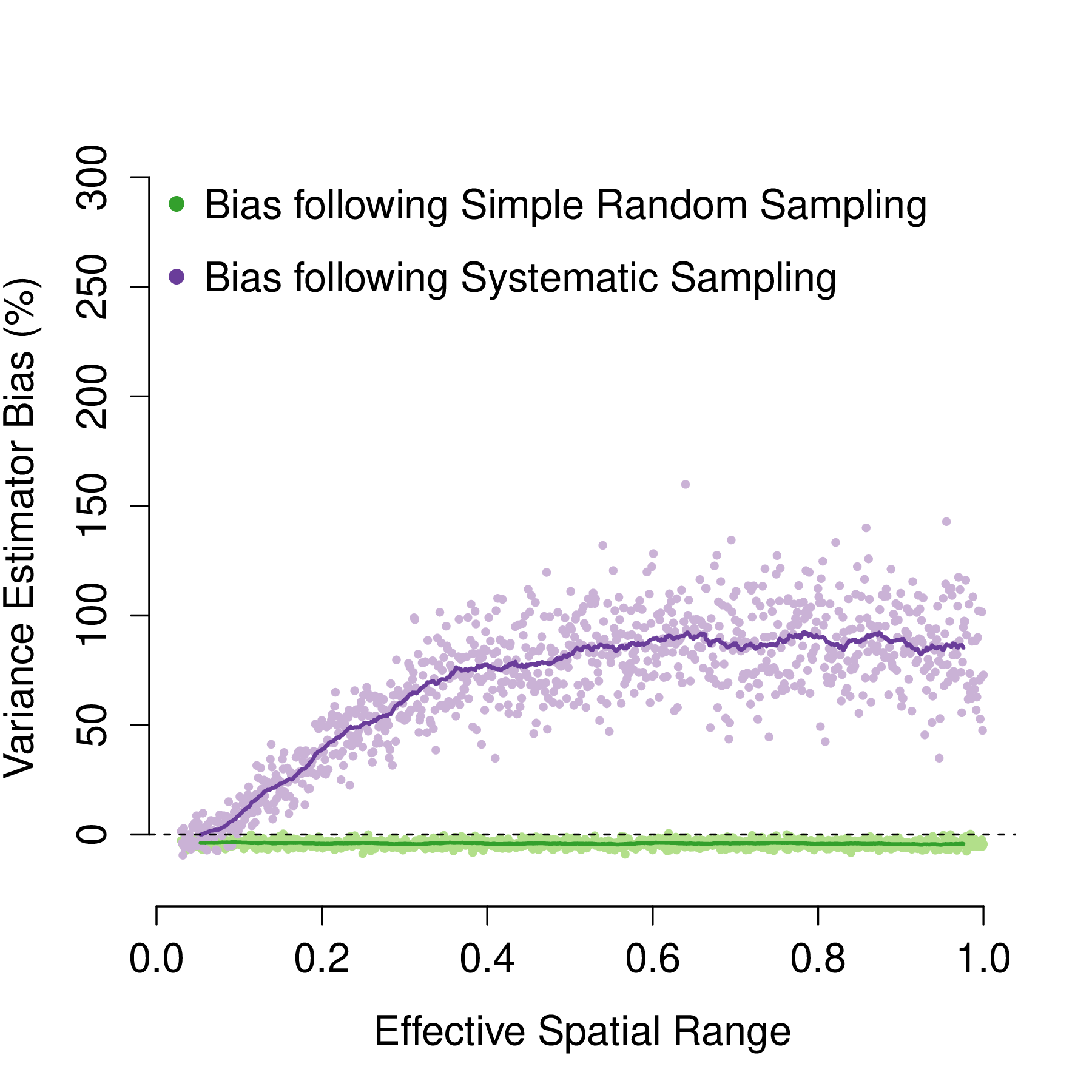}}
\subfloat[][\parbox{.2\textwidth}{GREG estimator 2 \\ sample size = 25}]{\includegraphics[width=.33\textwidth]{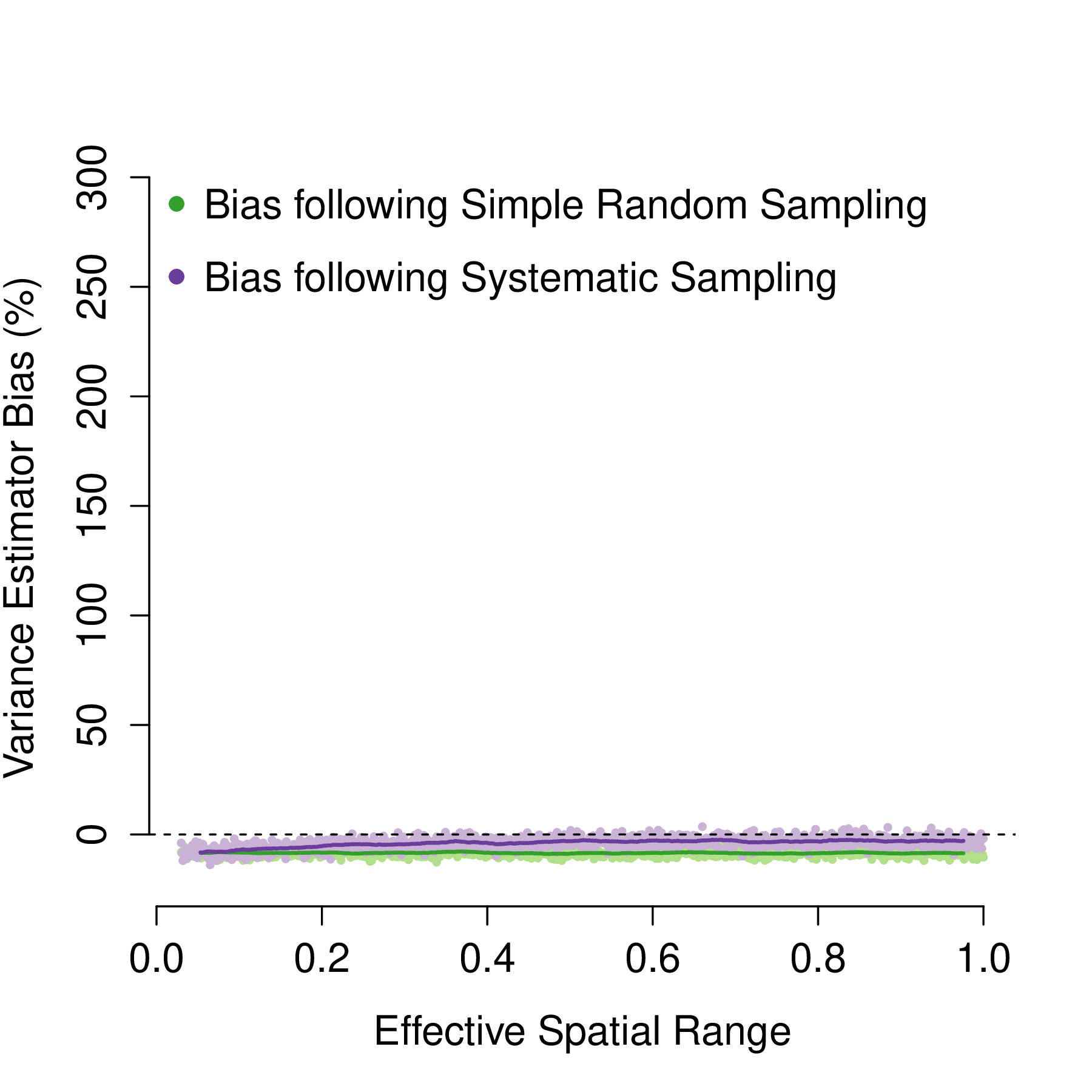}}\\
\subfloat[][\parbox{.2\textwidth}{H-T estimator \\ sample size = 100}]{\includegraphics[width=.33\textwidth]{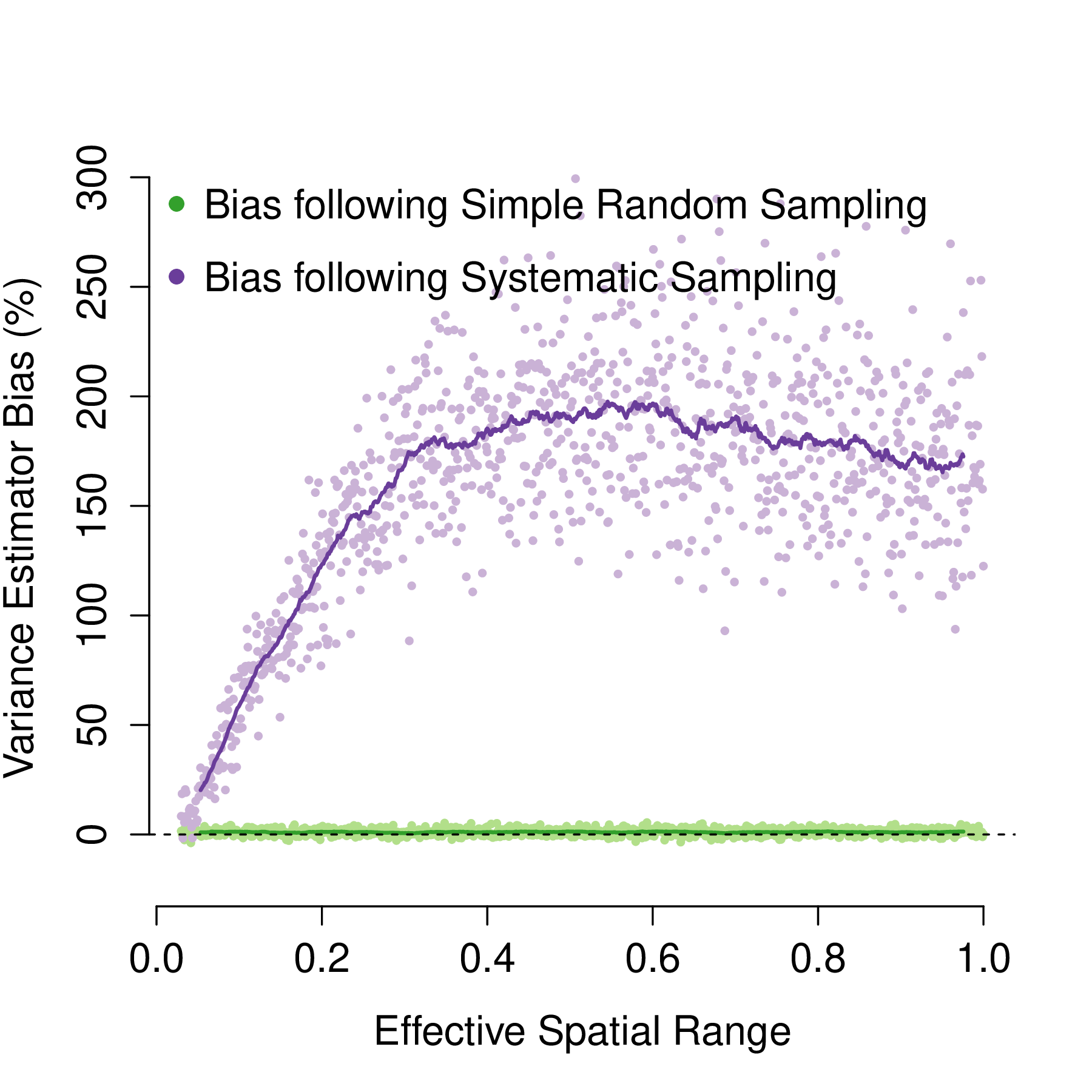}}
\subfloat[][\parbox{.2\textwidth}{GREG estimator 1 \\ sample size = 100}]{\includegraphics[width=.33\textwidth]{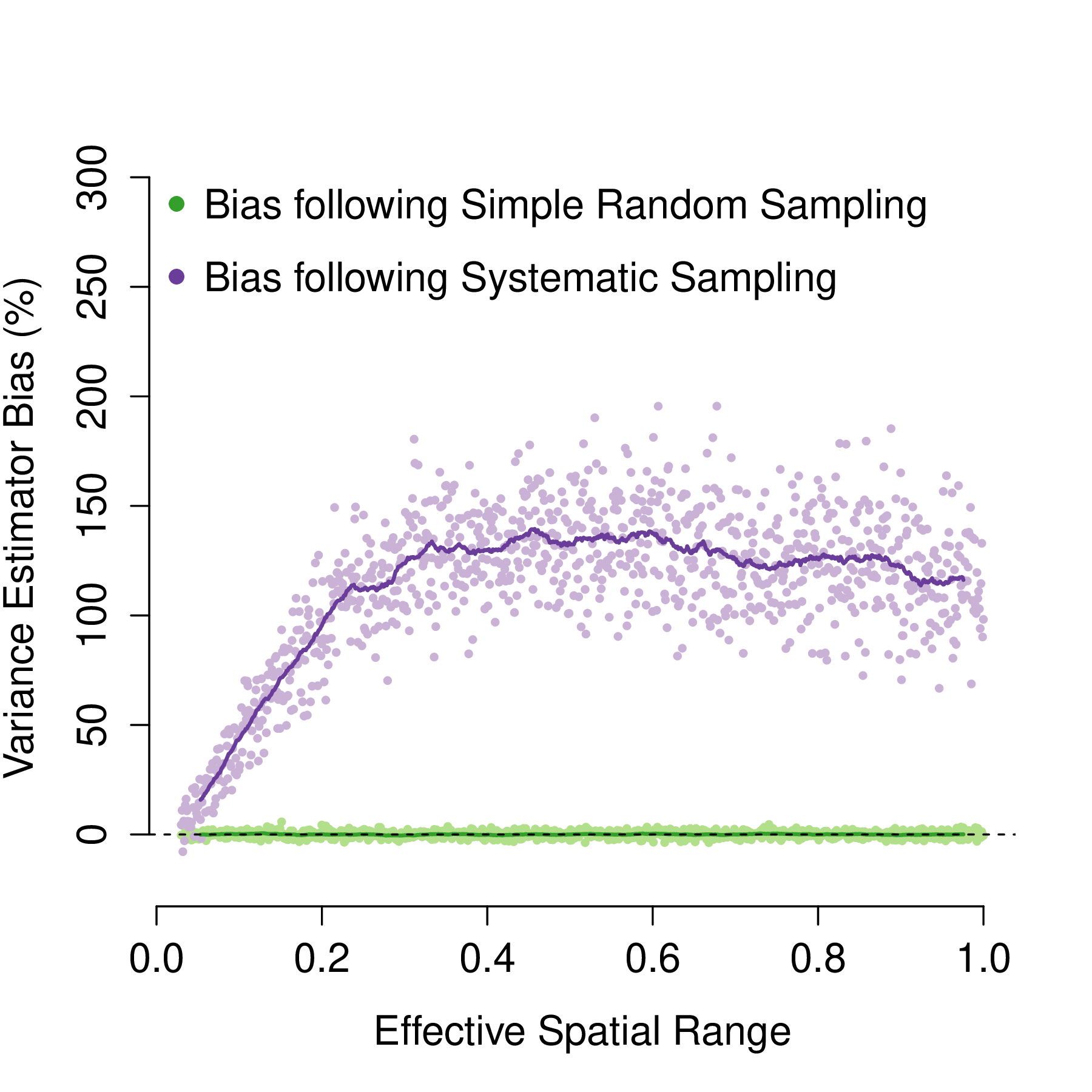}}
\subfloat[][\parbox{.2\textwidth}{GREG estimator 2 \\ sample size = 100}]{\includegraphics[width=.33\textwidth]{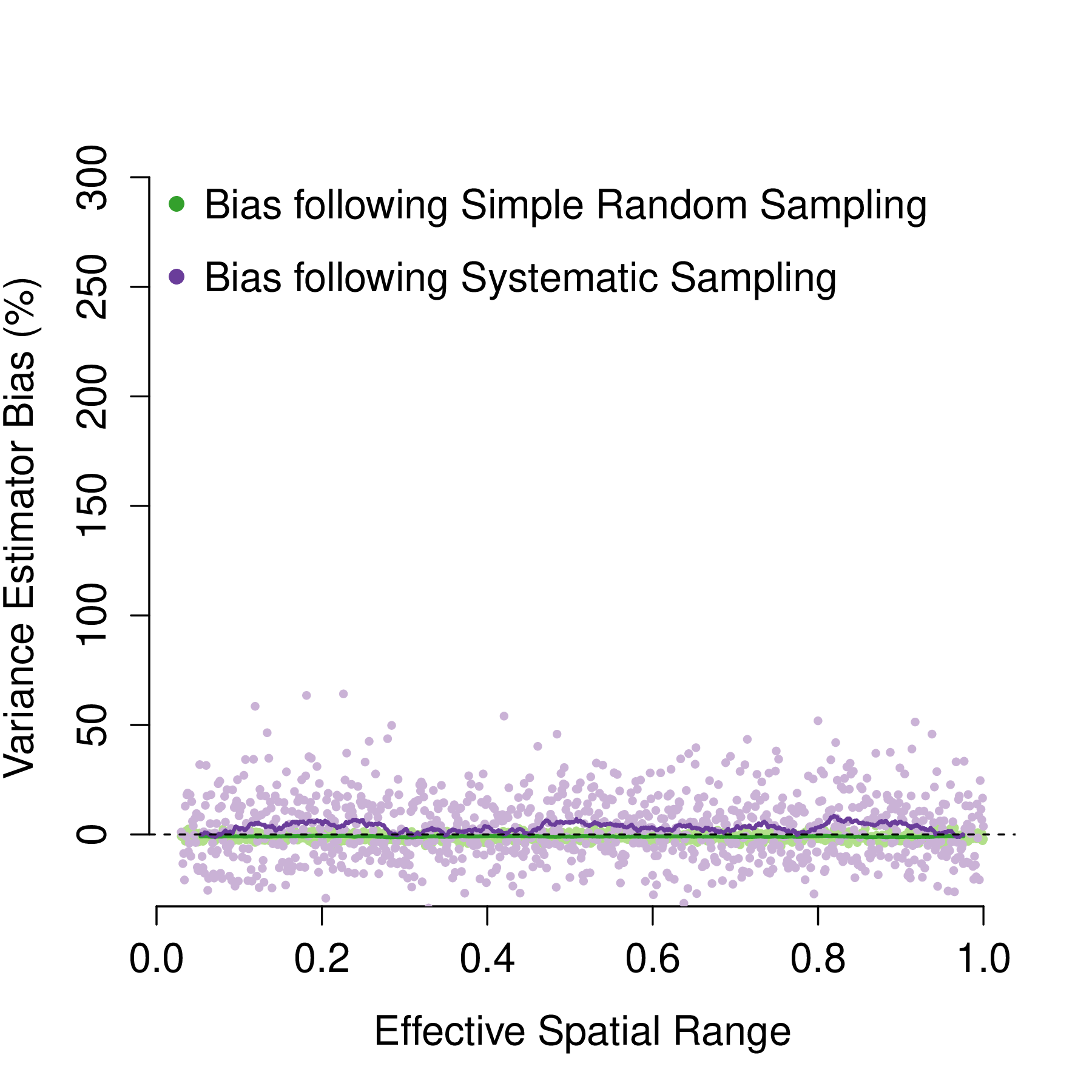}}
\caption{Comparisons between variance estimator bias following systematic and simple random sampling. Green and purple lines delineate the moving average of estimator bias following simple random sampling and systematic sampling, respectively. Horizontal gray dashed line is the zero bias reference line.}\label{synthetic-bias}
\end{figure}

Figure \ref{synthetic-bias} shows how variance estimator bias for the simple random and systematic sampling designs is affected by increasing spatial autocorrelation. The relative bias of each estimator of the variance of $\hat{\mu}$ was calculated as the mean estimated variance from the 1000 samples minus the empirical variance, expressed as a percentage of the empirical variance. We see that, for the systematic samples, the \emph{H-T estimator} and \emph{GREG estimator 1} tend conservative as spatial autocorrelation increases. For the systematic samples, bias increases quickly as \emph{esr} is increased from 0.03 to $\approx$ 0.5, after which an apparent asymptote is reached. Comparing Figure \ref{synthetic-bias}a with Figure \ref{synthetic-bias}d and Figure \ref{synthetic-bias}b with Figure \ref{synthetic-bias}e provides evidence that the apparent asymptote is higher for larger systematic sample sizes, indicating that increasing sample size may increase variance estimator percent bias for a particular population.

For the small sample size results ($n=25$), the GREG estimators show a slight negative bias following simple random sampling and following systematic sampling when residual spatial autocorrelation is not present, i.e., when population \emph{esr} is low for \emph{GREG estimator 1} (Figure \ref{synthetic-bias}b) and all synthetic populations for \emph{GREG estimator 2} (Figure \ref{synthetic-bias}c). It can also be seen that this bias becomes more negative as the number of covariates in the assisting model increases. This indicates that larger sample sizes may be required to estimate variances without bias when GREG estimators are used that leverage multiple covariates. Figures \ref{synthetic-bias}d, e and f indicate that this negative bias is not present for the larger sample size ($n=100$).

\subsection{Harvard Forest stem map analysis}
Table \ref{hf-tab} presents results for the Harvard Forest analysis. We see that the \emph{HF-H-T estimator} tends to over-estimate the variance following systematic sampling, evidenced by percent bias being $24.32\%$ and $57.47\%$ for the $n = 35$ and $140$ systematic sample sizes, respectively. We also see that variance estimator percent bias is greater for the larger sample size. This aligns with the synthetic populations analysis results that increasing sample size increases variance estimator percent bias for systematic sampling. Variance estimator bias is reduced when covariates are incorporated using the GREG estimators following systematic sampling, i.e., the \emph{HF-GREG estimator 1} and \emph{HF-GREG estimator 2} mean of estimated variances show a lower percent bias than the corresponding \emph{HF-H-T estimator} for each sample size. We further see that \emph{HF-GREG estimator 2} reduces variance estimator bias following systematic sampling more than \emph{HF-GREG estimator 1}. This result corroborates the results found in the synthetic populations analysis that reducing spatial autocorrelation in assisting model residuals reduces variance over-estimation (Figure \ref{varios} suggests that spatial autocorrelation in the residuals of \emph{HF-GREG estimator 2} $<$ \emph{HF-GREG estimator 1} $<$ \emph{HF-H-T estimator}).  

As with the synthetic populations analysis, \emph{HF-GREG estimator 1} and \emph{HF-GREG estimator 2} show a negative bias following simple random sampling for the small sample size ($n=35$). This corresponds with the synthetic populations analysis result that increasing the number of covariates used in the assisting model may require larger sample sizes for their associated variance estimators to be considered unbiased.

\begin{table}[!h]
\centering
\caption{Harvard Forest repeated sampling results. Values in parentheses indicate 95 percent bootstrap confidence intervals for empirical and mean of estimated variances.}\label{hf-tab}
\resizebox{\textwidth}{!}{ 
{\renewcommand{\arraystretch}{1.5}
   \begin{tabular}{@{}cc|ccc|ccc}\hline
                       &  & \multicolumn{3}{|c|}{n = 35} &  \multicolumn{3}{|c}{n = 140} \\\hline
     \multirow{2}{*}{
         \specialcell{Sampling\\[-.3em] design}
     }
     & \multirow{2}{*}{Estimator} & 
     \multirow{2}{*}{\specialcell{Mean of Estimated\\[-.3em] Variances}} & 
     \multirow{2}{*}{\specialcell{Empirical\\[-.3em] Variance}} & 
     \multirow{2}{*}{\specialcell{Percent\\[-.3em] Bias}} &
     \multirow{2}{*}{\specialcell{Mean of Estimated\\[-.3em] Variances}} & 
     \multirow{2}{*}{\specialcell{Empirical\\[-.3em] Variance}} & 
     \multirow{2}{*}{\specialcell{Percent\\[-.3em] Bias}}\\ &&&&&&&\\\hline
       \multirow{3}{*}{
         \specialcell{Simple\\[-.3em] Random}
     }
    &  HF-H-T estimator    & 1028 (1022, 1034) & 1026 (997, 1056) & 0.13\% &
                              257 (257, 258)   & 246 (239, 252)   & 4.86\%\\
    &  HF-GREG estimator 1 &  767 (763, 773)   & 773 (752, 795)   &-0.77\% & 
                              191 (190, 192)   & 187 (181, 192)   & 2.17\%\\
    &  HF-GREG estimator 2 &  702 (697, 706)   & 758 (736, 779)   &-7.36\% & 
                              176 (175, 177)   & 175 (170, 180)   & 0.57\%\\\hline
            \multirow{3}{*}{
         \specialcell{Systematic}
     }
     &  HF-H-T estimator    & 1032 (1027, 1038) & 830 (807, 853) & 24.32\% & 
                               257 (257, 258)   & 163 (159, 168) & 57.47\%\\
     &  HF-GREG estimator 1 &  772 (767, 777)   & 677 (658, 695) & 14.04\% & 
                               191 (190, 191)   & 145 (141, 149) & 31.84\%\\
     &  HF-GREG estimator 2 &  705 (701, 710)   & 678 (660, 697) &  3.98\% & 
                               175 (174, 176)   & 141 (138, 145) & 24.02\%\\\hline
    \end{tabular}}}
\end{table}

\clearpage

\section{Concluding Remarks and Recommendations}
The performance of the \emph{H-T estimator} on the simple random and systematic samples in the synthetic populations analysis provides evidence that increasing spatial autocorrelation increases systematic sampling efficiency. \citet[Example 3.4.2]{sarndal1992} demonstrates that ordering population units by their attribute value can increase within systematic sample heterogeneity thereby improving systematic sampling efficiency. Increasing spatial autocorrelation forces the attribute values for nearby population units to be more similar. This is comparable to how ordering population units by attribute value in a list forces neighbors to be more similar than if they were listed in random order. The systematic sampling designs employed here tend to produce samples with higher heterogeneity compared to simple random samples when spatial autocorrelation is increased because, as spatial autocorrelation increases, nearby population units become more similar.

Results from the synthetic populations and Harvard Forest analyses indicate that incorporating wall-to-wall remote sensing data can lower variance estimator bias following systematic sampling by reducing residual spatial autocorrelation. We see in the results figures and tables that stronger spatial autocorrelation in the population leads to greater over-estimation of variance for estimators that do not incorporate wall-to-wall ancillary data. However, when ancillary data that explains all or part of the spatial structure of the population is incorporated using a GREG estimator, variance estimator bias is reduced or eliminated. These results encourage the use of remote sensing information in forest inventory when systematic sampling is employed. For instance, the incorporation of readily available multispectral Landsat information could be used to curtail variance over-estimation that occurs when systematically placed cruise lines are used to conduct forest inventory.

Results show that, for the simple random sampling cases, there is a negative variance estimator bias for small sample sizes and that this bias becomes more negative as the number of assisting model covariates increases. This indicates that, when surveyors consider using complex assisting models to incorporate large amounts of ancillary remote sensing data in their estimation efforts, they need to consider increasing field sample sizes to account for increased negative variance estimator bias. 

This analysis only explored the use of wall-to-wall ancillary data to reduce variance estimator bias following systematic sampling. Often in large area inventories where the incorporation of remote sensing data is considered, only samples of remote sensing data are available \citep{andersen2011,babcock2018,ene2018,gregoire2011,saarela2015}. It is unclear whether the results obtained here apply directly to these more complex scenarios where multi-phase estimators need to be employed. In future research, the authors plan to examine the role of spatial autocorrelation in variance estimator bias using multi-phase estimators following systematic sampling.

\section{Acknowledgments}
The research presented in this study was partially supported by NASA's Arctic-Boreal Vulnerability Experiment (ABoVE) and Carbon Monitoring System (CMS) grants (proposal 13-CMS13-0006 funded via solicitation NNH13ZDA001N-CMS and proposal 15-CMS13-0012 funded via solicitation NNH15ZDA001N-CMS). Additional support was provided by the United States Forest Service Pacific Northwest Research Station. Andrew Finley was supported by National Science Foundation (NSF) DMS-1513481, EF-1137309, EF-1241874, and EF-1253225 grants.

\clearpage
\section*{Appendix}
Here we demonstrate that using the assisting models defined in Equations \ref{db-fit} and \ref{hf-db-fit} results in Equation \ref{ybar} reducing to the H-T estimator of the population mean following simple random sampling. We also demonstrate that Equation \ref{yse} becomes the H-T variance estimator following simple random sampling.\\

We note that $\hat{\beta}_0$ in Equations \ref{db-fit} and \ref{hf-db-fit} is obtained using ordinary least squares, i.e.,
\begin{equation}
\hat{\beta}_0 = (\bones^\top_n\bones_n)^{-1}\bones^\top_n\by_s = \frac{\bones^\top_n\by_s}{n}.
\end{equation}
Substituting the above result in Equations \ref{db-fit} and \ref{db-pred} (\ref{hf-db-fit} and \ref{hf-db-pred}) yields
\begin{align}
\hat{\by}_s & = \left(\frac{\bones^\top_n\by_s}{n}\right)\bones_n\qquad \text{and}\\
\by^\ast & = \left(\frac{\bones^\top_n\by_s}{n}\right)\bones_N,
\end{align}
which can be substituted into Equation \ref{ybar},
\begin{align}
\bar{y} &= \frac{\bones^\top_N\left(\frac{\bones^\top_n\by_s}{n}\right)\bones_N }{N} + 
\frac{\bones^\top_n\left(\left(\frac{\bones^\top_n\by_s}{n}\right)\bones_n - \by_s\right)}{n}\\
 &= \frac{\left(\frac{\bones^\top_n\by_s}{n}\right)\bones^\top_N\bones_N }{N} +
\frac{\left(\frac{\bones^\top_n\by_s}{n}\right)\bones^\top_n\bones_n}{n} - 
\frac{\bones^\top_n\by_s}{n}\\
&= \frac{\left(\frac{\bones^\top_n\by_s}{n}\right)N }{N} +
\frac{\left(\frac{\bones^\top_n\by_s}{n}\right)n}{n} - 
\frac{\bones^\top_n\by_s}{n}\\
&= \frac{\bones^\top_n\by_s}{n} +
\frac{\bones^\top_n\by_s}{n} - 
\frac{\bones^\top_n\by_s}{n}\\
&= \frac{\bones^\top_n\by_s}{n}\\
&= \frac{\sum_{i=1}^ny_{s_i}}{n},\label{ht-mean}
\end{align}
resulting in the H-T estimator for the population mean following simple random sampling. 

We note that $\hat{\by}_s = \bar{y}\bones_n$ and that only one parameter was estimated in Equations \ref{db-fit} and \ref{hf-db-fit} (i.e., $p = 1$). Substituting into Equation \ref{ys2} we obtain
\begin{align}
s^2 &= \frac{\bones_n^\top(\bar{y}\bones_n-\by_s)^2}{n-1}\\
&= \frac{\sum_{i=1}^n(\bar{y}-\by_{s_i})^2}{n-1}
\end{align}
which is the variance of $\by_s$. This leads to Equation \ref{yse} becoming the H-T variance estimator for (\ref{ht-mean}) following simple random sampling.

\clearpage
\newcommand{\bibliofont}{\footnotesize}

\end{document}